\documentclass[useAMS,usenatbib]{mnras}
\pdfoutput=0

\usepackage{amssymb}
\usepackage{amsmath}
\usepackage{times}
\usepackage{graphicx}
\usepackage{wasysym}

\def\aap{A\&A}
\def\apj{ApJ}

\def\apjl{ApJ}
\def\pasa{PASA}
\def\mnras{MNRAS}
\def\araa{ARA\&A}
\def\aj{AJ}

\def\apjs{ApJS}
\def\pasp{PASP}

\newcommand{\mincir}{\raise
-2.truept\hbox{\rlap{\hbox{$\sim$}}\raise5.truept 
\hbox{$<$}\ }}
\newcommand{\magcir}{\raise
-2.truept\hbox{\rlap{\hbox{$\sim$}}\raise5.truept
\hbox{$>$}\ }}
\newcommand{\minmag}{\raise-2.truept\hbox{\rlap{\hbox{$<$}}\raise
6.truept\hbox
{$>$}\ }}

\newcommand{\be}{\begin{equation}}
\newcommand{\ee}{\end{equation}}
\newcommand{\ba}{\begin{eqnarray}}
\newcommand{\ea}{\end{eqnarray}}
\newcommand{\brr}{\begin{array}}
 
\newcommand{\err}{\end{array}}
\newcommand{\bc}{\begin{center}}
\newcommand{\ec}{\end{center}}

\DeclareMathAlphabet{\mathsc}{OT1}{cmr}{m}{sc}
\def\testbx{bx}%
\DeclareRobustCommand{\ion}[2]{%
\relax\ifmmode
\ifx\testbx\f@series
{\mathbf{#1\,\mathsc{#2}}}\else
{\mathrm{#1\,\mathsc{#2}}}\fi
\else\textup{#1\,{\mdseries\textsc{#2}}}%
\fi}

\title[{\rm Exploring the quenching of galaxies with the sSFRF}]{The specific star formation rate function at different mass scales and quenching: A comparison between cosmological models and SDSS}

\author[A. Katsianis et al.]{Antonios Katsianis$^{1}$ $^{2}$ \thanks{E-mail:
     kata@sjtu.edu.cn}, Haojie Xu $^{2}$, Xiaohu Yang $^{1}$ $^{2}$,  Yu Luo $^{3}$, Weiguang Cui $^{4}$,  \newauthor Romeel Dav{\'e} $^{4}$, Claudia Del P. Lagos$^5$ $^6$ $^7$, Xianzhong Zheng  $^3$, Ping Zhao $^{2}$ \\
   \\ $^1$ Tsung-Dao Lee Institute, Shanghai Jiao Tong University, Shanghai 200240, China \\
   $^2$  Department of Astronomy, Shanghai Key Laboratory for Particle Physics and Cosmology, Shanghai Jiao Tong University, Shanghai 200240, China \\
   $^3$ Purple Mountain Observatory, No. 10 Yuanhua Road, Nanjing 210023, China \\
   $^4$ University of Edinburgh, Royal Observatory, EH9 3HJ Edinburgh, United Kingdom \\
    $^5$ International Centre for Radio Astronomy Research, University of Western Australia, 35 Stirling Hwy, Crawley, WA 6009, Australia \\
   $^6$ Cosmic Dawn Center (DAWN), Denmark, Norregade 10, DK-1165 Kobenhavn, Denmark\\
   $^7$ ARC Centre of Excellence for All Sky Astrophysics in 3 Dimensions (ASTRO 3D) \\
   } 

\begin{document}

\maketitle

\begin{abstract}

  We present the eddington bias corrected Specific Star Formation Rate Function (sSFRF) at different stellar mass scales from a sub-sample of the Sloan Digital Sky Survey Data Release DR7 (SDSS), which is considered complete both in terms of stellar mass (${\rm M_{\star}}$) and star formation rate (SFR). The above enable us to study qualitatively and quantitatively quenching, the distribution of passive/star-forming galaxies and perform comparisons with the predictions from state-of-the-art cosmological models, within the same ${\rm M_{\star}}$ and SFR limits. We find that at the low mass end (${\rm M_{\star}} = 10^{9.5} - 10^{10} \, {\rm M_{\odot}}$) the sSFRF is mostly dominated by star-forming objects. However, moving to the two more massive bins (${\rm M_{\star}}  = 10^{10} - 10^{10.5} \, {\rm M_{\odot}}$ and ${\rm M_{\star}} = 10^{10.5} - 10^{11} \, {\rm M_{\odot}}$) a bi-modality with two peaks emerges. One peak represents the star-forming population, while the other describes a rising passive population. The bi-modal form of the sSFRFs is not reproduced by a range of cosmological simulations  (e.g. Illustris, EAGLE, Mufasa, IllustrisTNG) which instead generate mostly the star-forming population, while a bi-modality emerges in others (e.g. L-Galaxies, Shark, Simba). Our findings reflect the need for the employed quenching schemes in state-of-the-art models to be reconsidered, involving prescriptions that allow ``quenched galaxies'' to retain a small level of SF activity (sSFR $=$ ${\rm 10^{-11} {\rm yr^{-1}}}$-${\rm 10^{-12} {\rm yr^{-1}}}$) and generate an adequate passive population/bi-modality even at intermediate masses (${\rm M_{\star}}  = 10^{10} - 10^{10.5} \, {\rm M_{\odot}}$). 

\end{abstract}

\begin{keywords}
cosmology: theory -- galaxies: formation -- galaxies: evolution -- methods: numerical
\end{keywords}

\section{Introduction}
\label{intro}

In the last decade the observed star formation rates (SFRs) and stellar masses (${\rm M_{\star}}$) of galaxies have been extensively used to constrain and test theoretical models of galaxy formation. Cosmological simulations such as ANGUS \citep{TescariKaW2013,Katsianis2014,Garcia2017}, Illustris \citep{Vogelsberger2014}, Horizon-AGN \citep{Dubois2014,Volonteri2016}, EAGLE \citep{Schaye2015,Crain2015},  Mufasa \citep{Dave2017}, IllustrisTNG  \citep{Pillepich2018} and SIMBA \citep{dave2019} have used sub-grid  models to reproduce realistic galaxies in terms of stellar mass and SFR \citep[]{Katsianis2016,Dave2017,Zhaoka2020}. In addition, semi-analytic models of galaxy formation such as L-GALAXIES \citep{Henriques2015,Luo16} and Shark \citep{Shark2019,Davies2019} have given us the opportunity to study galaxy formation in larger volumes and be tested against stellar masses and SFRs at multiple redshifts. A useful metric, widely used in the literature to study both star formation and its efficiency is the Specific Star Formation Rate (sSFR) defined as $ {\rm sSFR = SFR/M_{\star}}$. Hence, some authors have explored the number density of galaxies in bins of sSFRs namely the Specific Star Formation Rate Function \citep[sSFRF, ][]{Ilbert2015,Dave2017,Katsianis2019}. The sSFRF has advantages against median/average relations in 2d scatter plots which are not able to provide quantitative information of how galaxies are distributed and do not account for galaxies that could be under-sampled or missed by selection effects \citep{Ilbert2015}.  

In order to study quenching in galaxies, it has been a common practice to study objects and their number distributions in terms of optical colors \citep{Strateva2001}, but a severe complication is that the latter strongly depend on the properties of the stellar population (e.g. metallicities, age distribution) and dust attenuation \citep{Cantiello2007,Li2007a,Carter2009,Trayford2017,Bravo2020}. This makes colors a less direct probe of the star formation activity of an object. Thus, in the last decade different studies have instead relied on the sSFR as a more direct probe of galaxy quenching \citep{Darvish2018,Belfiore2018,Davies2019b}. Furthermore, sSFRs have the advantage of being more easily comparable to the predictions of cosmological simulations.

When all galaxies are considered together (regardless of photometry, spectroscopy or morphology), the color distribution of galaxies can be approximated by the sum of two Gaussian functions, i.e. a bi-modal function \citep{Baldry2008} with one representing star-forming objects and the other quenched galaxies. The sSFR is a property that directly relates with optical color, besides degeneracies due to dust attenuation, age and metallicity, so some authors expect or have found a similar color distribution as well. \citet{Santini2009} demonstrated that the sSFR of their sample from the GOODS-MUSIC catalog at ${\rm z \sim 0.3-2.5}$ shows a well-defined bi-modal distribution, with a clear separation between the actively star-forming objects and passive galaxies. In addition, \citet{Tzanavaris2010} showed a clear bi-modal sSFR distribution in their sample compiling UV data from the Swift UV/ Optical Telescope and mid-IR data from the Spitzer SpaceTelescope MIPS 24$\mu$m camera. \citet{Wetzel2011} using catalogs created from the SDSS survey \citep{Abazajian2009}, examined the sSFR distribution of satellite  galaxies  and  its  dependence  on  stellar  mass,  host  halo  mass,  and  halo-centric radius. The authors demonstrated that all galaxies, regardless of being central or satellites, exhibit a similar bi-modal sSFR distribution, with satellites being more likely to belong to the quenched population. \citet{Lenkic2016} suggested that the sSFRs in their sample constructed with Swift, Spitzer IRAC and MIPS 24 ${\rm \mu m}$ photometry of 183 galaxies in 46 compact groups, separate into two populations with one corresponding once again to the star-forming objects while the other to the quiescent galaxies. The authors demonstrated that galaxies in HI-rich groups tend to be MIR active, UV blue and have on average higher sSFRs, while galaxies residing in HI-poor groups populate the ``low end of the sSFR distribution''. \citet{dave2019} demonstrated that a bi-modal form emerges for the histograms of observed sSFRs obtained from the GALEX-SDSS-WISE  Legacy  CatalogS \citep[GSWLC, ][]{Salim2016} and the Simba simulations.

In this work we use the Sloan Digital Sky Survey Data Release (SDSS) to construct specific Star Formation Rate Functions (sSFRFs) within different mass scales (section \ref{SFRSSDSS}) from a sub-sample which is considered to have robust SFRs and stellar masses (has been compared successfully with other studies and SFR indicators) following \citet{Zhaoka2020}. The objective is to study qualitatively the number density/distribution and quenching across different mass scales within strict SFRs and $\rm M_{\star}$ limits. We compare the observational constraints with the predictions from state-of-the-art cosmological models (section \ref{simsbina}). In section \ref{con} we present our conclusions.

\begin{figure*}
\centering
\includegraphics[scale=0.47]{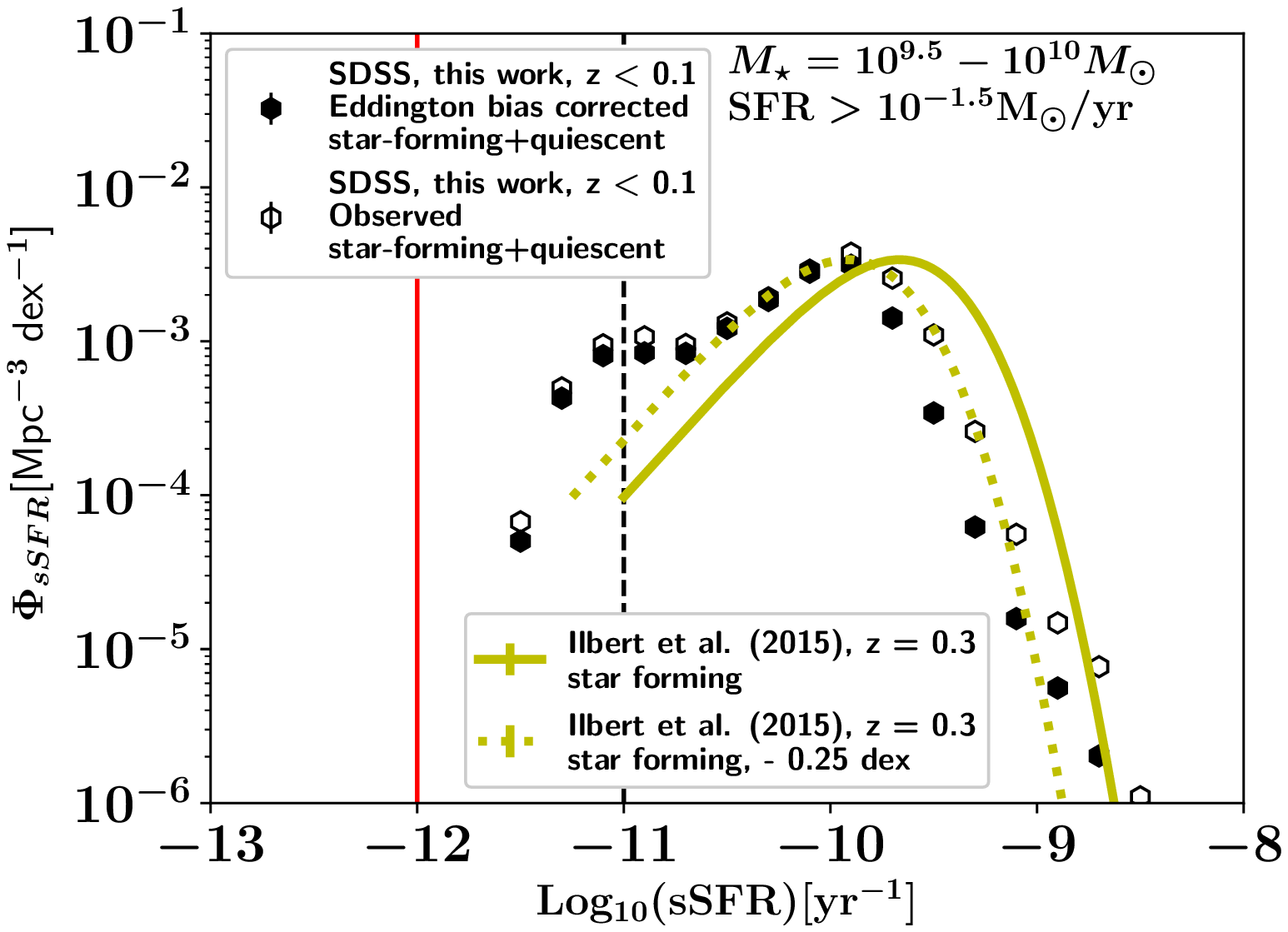}
\includegraphics[scale=0.47]{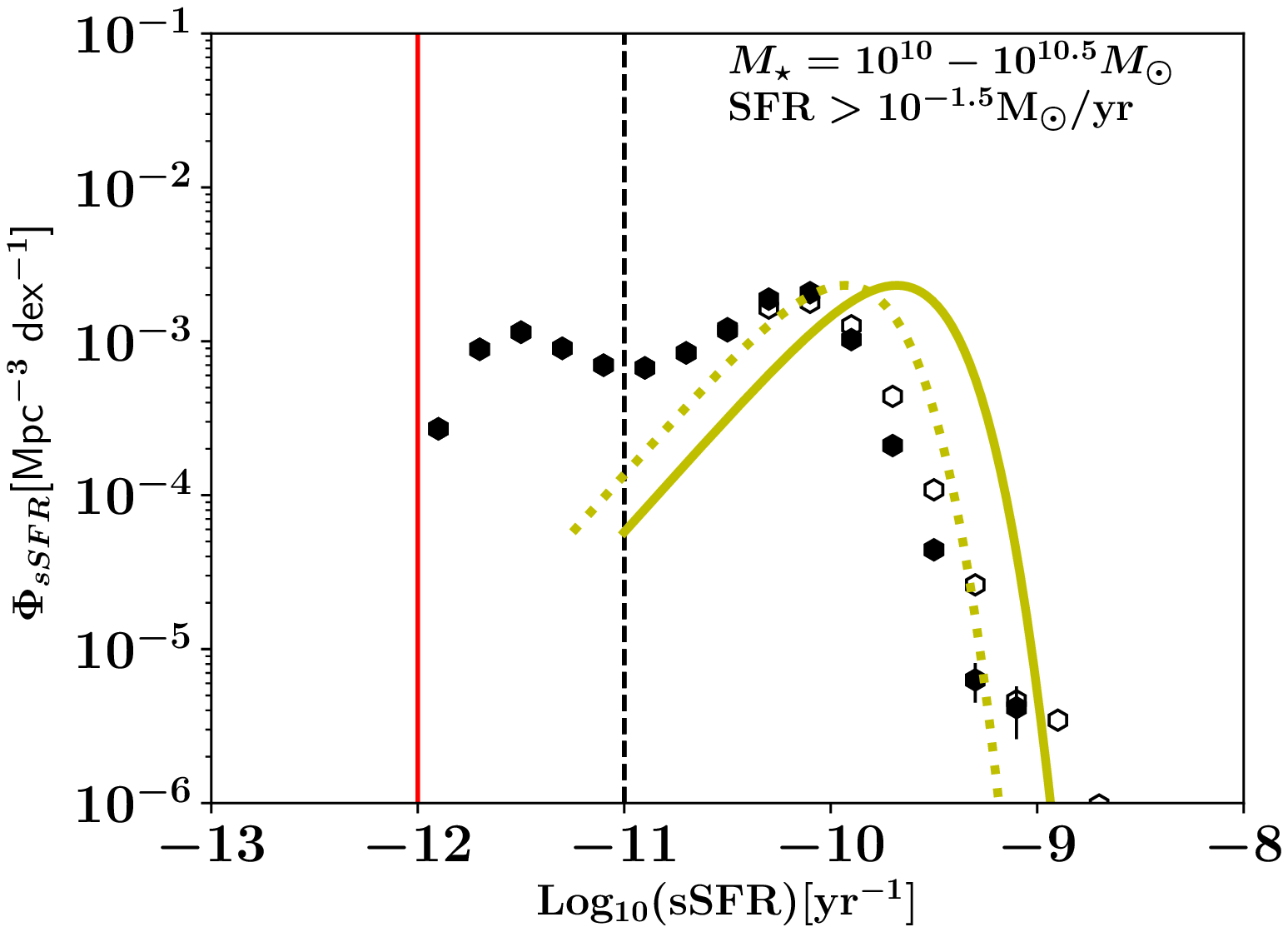} \\
\vspace{-0.76cm}
\includegraphics[scale=0.47]{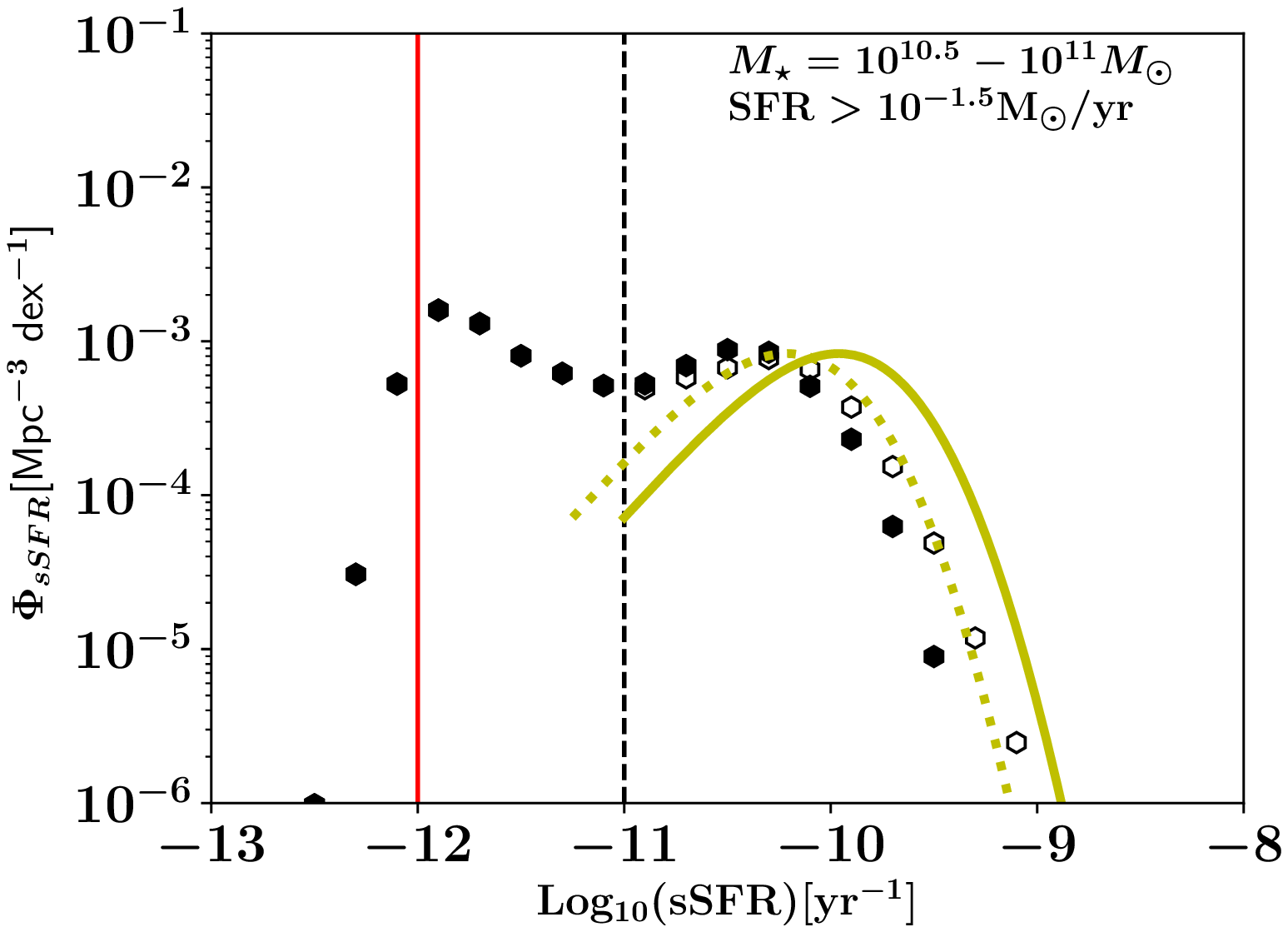}
\includegraphics[scale=0.47]{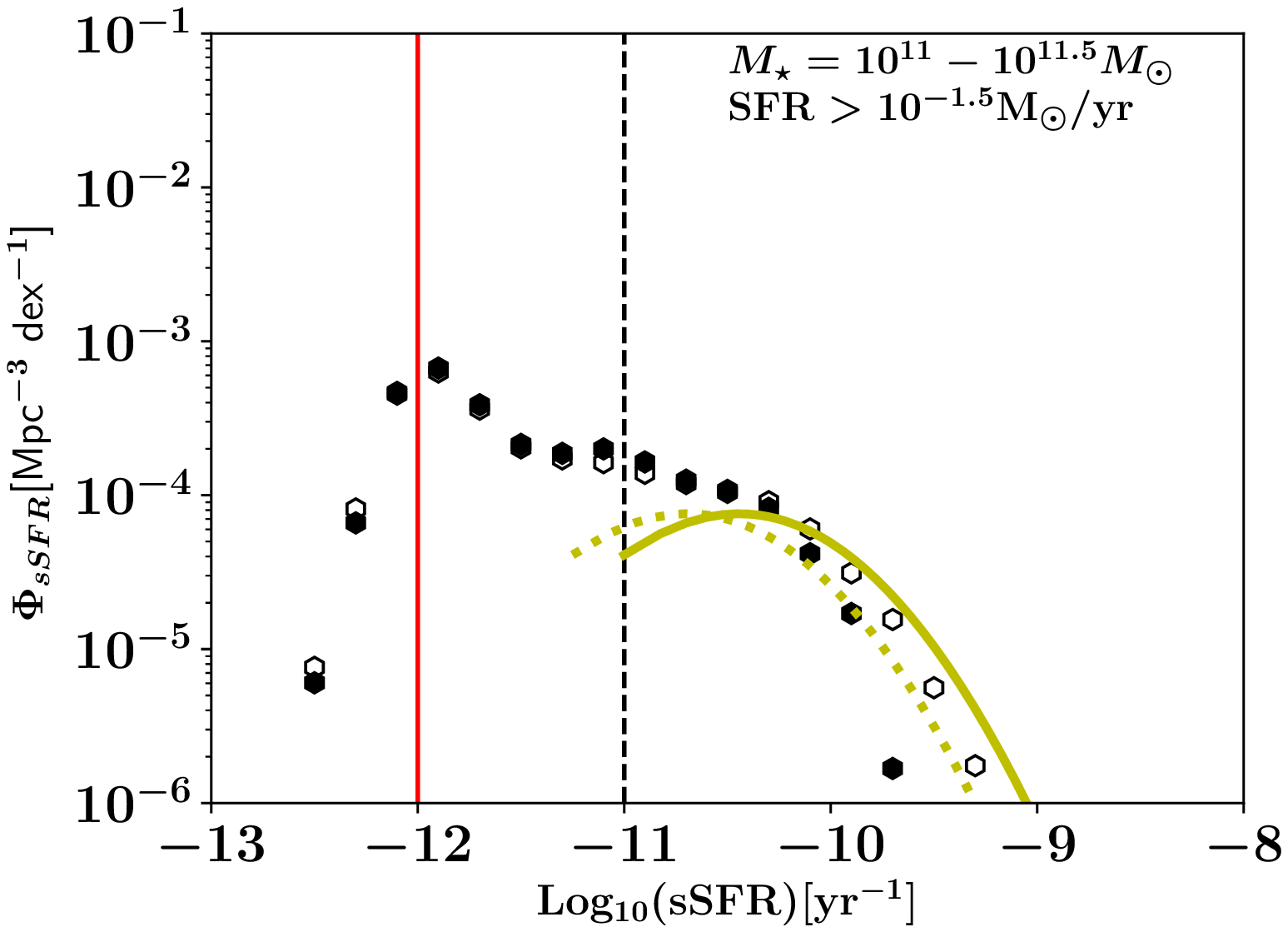}
\caption{The black solid hexagons represent the Eddington bias corrected specific star formation rate function (sSFRF) at ${\rm z \sim 0}$ from the SDSS at mass intervals of $10^{9.5} - 10^{10.0} \, {\rm M_{\odot}}$ (top left panel), $10^{10.0} - 10^{10.5} \, {\rm M_{\odot}}$ (top right panel), $10^{10.5} - 10^{11.0} \, {\rm M_{\odot}}$ (bottom left panel) and $10^{11.0} - 10^{11.5} \, {\rm M_{\odot}}$ (bottom right panel). Open symbols represent the observed biased sSFRF. The black vertical line marks the limit of sSFR $=$ $10^{-11} {\rm yr^{-1}}$ with galaxies having lower values labelled as passive/quenched objects and galaxies having higher values labelled as star forming. For comparison we provide the observed sSFRs of \citet{Ilbert2015} which include only star forming objects, defined by a main sequence using a color-color selection) represented by the orange solid line and orange dotted line (shifted by -0.25 dex in order to consider systematics between star formation rate indicators). SDSS sSFRs  $\le 10^{-12} \, {\rm yr^{-1}}$ should only be considered as upper limits to the true values and we visualize this regime via the red vertical line.}
\label{fig:sSFRIR1}
\end{figure*}

\section{The specific star formation rate function in SDSS at different mass scales}
\label{SFRSSDSS}

\begin{table*}
  \centering
\resizebox{0.90\textwidth}{!}{%
  \begin{tabular}{cccccc}
    \hline \\
    & {\large ${\rm Log_{10}(sSFR)}$ (${\rm yr^{-1}})$} & & 
    {\large  $ \phi_{\rm sSFR}\ \left({\rm Mpc}^{-3} \, dex^{-1}\ \right) $ } \\ \\
    \hline \hline
    &  & $10^{\, 9.5-10}$ ${\rm M_{\odot}}$ & $10^{\, 10-10.5}$ ${\rm M_{\odot}}$  & $10^{\, 10.5-11}$  ${\rm M_{\odot}}$ &  $10^{\, 11-11.5}$ ${\rm M_{\odot}}$\\ 
    \hline
    &-12.900000 & - & - & - &  1.38E-08 $\pm$ 1.17E-08 \\
    &-12.700000 & - & - & - &  4.84E-07 $\pm$ 1.23E-07 \\
    & -12.500000 &  -  & -  & 9.75E-07 $\pm$ 3.25E-07  & 6.04E-06 $\pm$ 5.94E-07 \\
  &  -12.300000  & - & -  & 3.06E-05 $\pm$ 2.02E-06  & 6.60E-05 $\pm$ 2.41E-06 \\
& -12.100000 & - & - & 5.29E-04 $\pm$ 1.74E-05 & 4.53E-04 $\pm$ 1.10E-05 \\
& -11.900000 & - & 2.69E-04 $\pm$ 1.31E-05  & 1.59E-03 $\pm$ 5.01E-05 & 6.69E-04 $\pm$ 1.67E-05 \\ 
& -11.700000 &  - & 8.87E-04 $\pm$ 4.04E-05 & 1.31E-03 $\pm$ 3.81E-05 & 3.86E-04 $\pm$ 1.10E-05 \\
& -11.500000 & 5.02E-05 $\pm$ 6.99E-06  & 1.15E-03 $\pm$ 5.47E-05  & 8.05E-04 $\pm$ 2.44E-05  & 2.13E-04 $\pm$ 6.11E-06 \\
& -11.300000 & 4.27E-04 $\pm$ 4.06E-05 & 8.99E-04 $\pm$ 4.51E-05 & 6.19E-04 $\pm$  1.85E-05 &  1.86E-04 $\pm$ 5.41E-06 \\ 
& -11.100000 & 8.06E-04 $\pm$ 7.00E-05 & 6.99E-04 $\pm$ 3.45E-05  & 5.16E-04 $\pm$ 1.54E-05  &  1.99E-04 $\pm$ 5.34E-06 \\ 
& -10.900000 & 8.40E-04 $\pm$ 7.38E-05 & 6.71E-04 $\pm$ 2.96E-05 &  5.28E-04 $\pm$ 1.38E-05  &  1.64E-04 $\pm$ 4.83E-06 \\ 
& -10.700000 & 8.34E-04 $\pm$ 5.31E-05 & 8.40E-04 $\pm$ 3.23E-05 & 6.94E-04  $\pm$ 1.81E-05 &  1.25E-04 $\pm$ 3.46E-06 \\
& -10.500000 & 1.21E-03 $\pm$ 7.88E-05 & 1.20E-03 $\pm$ 3.94E-05 & 8.83E-04 $\pm$  2.24E-05 &  1.04E-04 $\pm$ 3.08E-06 \\
& -10.300000 & 1.91E-03  $\pm$ 9.66E-05 & 1.88E-03 $\pm$ 5.88E-05 & 8.48E-04 $\pm$ 2.30E-05 &  8.21E-05 $\pm$ 2.87E-06 \\
& -10.100000 & 2.81E-03 $\pm$ 1.29E-04 & 2.07E-03 $\pm$ 7.18E-05 &  5.09E-04 $\pm$ 1.45E-05 &  4.16E-05 $\pm$ 1.63E-06 \\
& -9.900000 & 3.11E-03 $\pm$ 1.28E-04 & 1.03E-03 $\pm$ 3.95E-05  & 2.31E-04 $\pm$ 8.25E-06 &  1.70E-05 $\pm$ 9.72E-07  \\
& -9.700000 & 1.41E-03 $\pm$ 8.18E-05 & 2.10E-04 $\pm$ 1.19E-05  & 6.26E-05 $\pm$ 3.09E-06 & 1.67E-06 $\pm$ 2.83E-07  \\
& -9.500000 & 3.42E-04 $\pm$ 3.68E-05  & 4.42E-05 $\pm$ 4.64E-06 & 8.94E-06 $\pm$ 1.28E-06 &  4.26E-08 $\pm$ 4.26E-08 \\
& -9.300000 & 6.20E-05 $\pm$ 1.21E-05 & 6.30E-06 $\pm$ 1.87E-06  & 7.35E-07 $\pm$ 3.85E-07 & - \\
& -9.100000 & 1.57E-05 $\pm$ 7.74E-06 &  4.16E-06 $\pm$ 1.57E-06 & - & - \\
& - 8.900000 & 5.58E-06 $\pm$ 3.37E-06 & 4.54E-07 $\pm$ 4.54E-07 & - & - \\
\hline \hline
  \end{tabular}%
}
\caption{The Specific Star Formation Rate Function in the SDSS at mass bins of ${\rm M_{\star}} = 10^{\, 9.5-10}$ ${\rm M_{\odot}}$ (second column), ${\rm M_{\star}} = 10^{\, 10-10.5}$ ${\rm M_{\odot}}$ (third column), ${\rm M_{\star}} = 10^{\, 10.5-11}$ ${\rm M_{\odot}}$ (forth column) and ${\rm M_{\star}} = 10^{\, 10.5-11}$ ${\rm M_{\odot}}$ (fifth column).}
\label{tab:sim_runs}
\end{table*}

The Sloan Digital Sky Survey Data Release 7 \citep[SDSS DR7][]{Abazajian2009,Blanc2019,ConcasA2019,Zhaoka2020,LuYi2020} is one of the most well-studied galaxy surveys at ${\rm z \sim 0}$. It provides stellar masses and SFRs for more than  500,000 galaxies which are based on the spectral distributions  of \citet{Brinchann04}, prescriptions  for  active  galactic  nuclei contamination and fiber aperture corrections following \citet{Salim2007}. The SFRs  are inferred from the Spectral Energy Distribution (SED) fitting of emission lines (predominately H${\rm \alpha}$), but in the cases of strong AGN contamination or no measurable emission lines, SFRs are obtained primarily using the 4000 ${\rm \AA}$ break. The stellar masses and SFRs described above have been used to estimate the local Galaxy Stellar Mass Function \citep[GSMF, ][]{Baldry2008} and Star Formation Rate Function \citep[SFRF, ][]{Zhaoka2020}, respectively. The resulting distributions have been in agreement with a range of studies that employ different SFR indicators regardless of any shortcomings of SDSS. The GSMF and SFRF have been found to have confidence limits of $10^{9} $ ${\rm M_{\odot}}$ in terms of stellar mass \citep{Weigel2016} and $10^{-1.5}$ ${\rm M_{\odot} / {\rm yr}}$ in terms of SFR \citep{Zhaoka2020} so for our analysis we consider only objects within these strict ranges to construct the sSFRF. Given our adopted ${\rm M_{\star}}$ and SFR confidence limits, our sSFRFs are formally complete above ${\rm Log_{10}(sSFR)} > 10^{-10.5}$ ${\rm yr^{-1}}$. Any incompleteness below these limits is coming from sources that are not present/observed in the SDSS main catalog. Thus, we caution the reader that any comparisons with the predictions from cosmological models should be done within the same ${\rm M_{\star}}$ and SFR limits. For this work we assume a \citet{chabrier03} initial mass function and a \citet{Planck2016} cosmology. We convert the stellar masses and SFRs from \citet{Kroupa} to \citet{chabrier03} by multiplying them by a factor of 1.063 following \citet{Bell2003} and \citet{Madau2014}. For every galaxy, the sSFR is the SFR/${\rm M_{\star}}$ ratio so the initial values are unchanged. When necessary we shift the cosmology of other observations or simulations to \citet{Planck2016} following \citet{Croton2013}.

We measure the sSFRFs in volume-limited samples that have been found complete in stellar mass and r-band luminosity \citep{Vadenbosch2008}. In order to correct for the Malmquist bias caused by the nature of flux limit of the survey, we start with the observed sSFRFs computed via the 1/Vmax method \citep{Li2009}.  The Vmax is calculated from the r-band Petrosian magnitude (K+E corrected to z=0.1), with also spectroscopy completeness taken into account. The observed distribution, due to the error on the sSFR estimation, suffers from the \citet{Eddington1913} bias which we correct following \citet{Zhaoka2020}. In short, without  resort  to  assuming  a  functional  form  for  the  intrinsic  (Eddington  bias  corrected) SFRFs and galaxy stellar mass functions, we correct the Eddington bias in both by subtracting the SFR/${\rm M_{\star}}$ of each galaxy using the average shift in the distributions induced by the bias. Then we employ the required corrections to the sSFRF. We start by convolving the SFRFs and GSMFs with SFR/${\rm M_{\star}}$ uncertainties taken from the assumption that each galaxy follows a Gaussian distribution around its median SFR/${\rm M_{\star}}$ and the uncertainty is the standard deviation. We obtain 1000 SFRs/${\rm M_{\star}}$ for each galaxy and then we build a histogram from the 1000 mock distributions (the median value of the resulting mocks). By doing this, we mimic the Eddington bias (EB) effect and create a double EB contaminated distribution. After comparing the latter with the observed SFRF in the x-axis, we interpolate and apply the necessary correction to the SFR of each galaxy in order to remove the Eddington bias. We then build another histogram which we convolve to generate a first order correction for the SFRF. We then again convolve the latter with the SFR uncertainties to obtain a first level Eddington bias corrected SFRF. If the latter distribution match the observed SFRF, it is implied that the above process should be enough to produce an intrinsic Eddington bias free SFRF. If not then we repeat the procedure by applying an additional correction N times, until the N-th order Eddingthon bias corrected SFRF converges to the observed distribution. For the SFRF and GSMF from the SDSS data only 2 iterations were necessary. The intrinsic SFRF (EB corrected) can then be found by applying these steps to the observed SFRFs/GSMFs. We then perform the corrections seen for the SFRF and GSMF to the specific star formation rates to obtain an Eddington bias corrected sSFRF.

The black solid hexagons of Fig. \ref{fig:sSFRIR1} represent the sSFRF at $z \sim 0$ from SDSS DR 7 at the stellar mass intervals of ${\rm M_{\star}} = 10^{9.5} - 10^{10.0} \, {\rm M_{\odot}}$ (top left panel), ${\rm M_{\star}} = 10^{10.0} - 10^{10.5} \, {\rm M_{\odot}}$ (top right panel), ${\rm M_{\star}} = 10^{10.5} - 10^{11.0} \, {\rm M_{\odot}}$ (bottom left panel) and ${\rm M_{\star}} = 10^{11.0} - 10^{11.5} \, {\rm M_{\odot}}$ (bottom right panel). The open symbols represent the observed Eddington biased sSFRF while the black vertical line labels the sSFR $=$ ${\rm 10^{-11} {\rm yr^{-1}}}$ limit which separates the passive/quenched objects and star forming objects. The above limit to separate the two populations is commonly adopted in the literature \citep{Cassata2010,Tamburri2014,Katsianis2019,Matthee2018,ThomasR2019,Lovell2020} and is equivalent to the selection made by the NUV-r-J diagram \citet{Ilbert2013}. At the same time it has the advantage of relating directly the separation of populations using a sSFR cut instead. We note that the applied Eddington bias correction does not change the form of the sSFRF. This  is  probably  due  to  the  cancellation  of  the  Eddington  bias that  affects similarly both the star formation rate function and stellar mass function.

Starting from the low mass interval of ${\rm M_{\star}} = 10^{9.5} - 10^{10.0} \, {\rm M_{\odot}} $ we see that there is a peak of the distribution for objects with sSFRs of $10^{-10} {\rm yr^{-1}}$ pointing to the direction that most of the low-mass galaxies in our sample are still star forming. In addition, there is a second smaller peak with sSFR = $10^{-11} {\rm yr^{-1}}$. Moving to the mass interval of ${\rm M_{\star}} = 10^{10.0} - 10^{10.5} \, {\rm M_{\odot}}$ we demonstrate that the shape of the sSFR is bi-modal with two peaks at sSFR = ${\rm 10^{-10.2} {\rm yr^{-1}}}$ (representing star-forming galaxies) and sSFR = ${\rm 10^{-11.5} {\rm yr^{-1}}}$ (quiescent objects) suggesting that at these intermediate stellar masses there are numerous passive objects. At the more massive interval of ${\rm M_{\star}} = 10^{10.5} - 10^{11} \, {\rm M_{\odot}}$ (which represents objects with masses close to the characteristic stellar masses of the stellar mass function) we encounter once again a bi-modal form for the sSFRF, while the peak that represents quiescent objects (found at sSFR = ${\rm 10^{-11.9} {\rm yr^{-1}}}$) is slightly larger than that of star-forming galaxies (found at sSFR = ${\rm 10^{-10.5} {\rm yr^{-1}}}$). We note that both the star-forming and quenched peaks move towards lower sSFRs suggesting that as we move towards larger masses galaxies become more passive as a whole. High-mass galaxies tend to accumulate their stellar mass earlier than low-mass galaxies and thus on average be more passive \citep{Fang2018}. The above trend, called the sSFR downsizing phenomenon, has been widely seen in the literature \citep{Cowie2008,Firmani2010,Twite2011,Hall2018}. Last, at the high mass end (${\rm M_{\star}} = 10^{11.0} - 10^{11.5} \, {\rm M_{\odot}}$) we see the presence of a considerable star-forming population but most galaxies are quenched with the sSFRF having a peak at sSFR = ${\rm 10^{-11.9} {\rm yr^{-1}}}$.

We note that \citet{Wetzel2011} besides concluding that the distribution of sSFRs is bi-modal in SDSS argued that the high value and sharp peak of the low sSFRs at sSFR = ${\rm 10^{-12} {\rm yr^{-1}}}$ in their study,  is partially driven by the limitation that low sSFR galaxies  with  no  detectable emission lines are assigned SFRs mostly based on ${\rm D_{4000}}$. These are considered uncertain and usually upper limits \citep{Feldmann2017} of the true values (i.e. there are a lot of objects that have artificial sSFRs $\sim $ ${\rm 10^{-12} {\rm yr^{-1}}}$). In addition, \citet{Hahn2018} emphasized that SDSS sSFRs  $\le 10^{-12} \, {\rm yr^{-1}}$ should only be considered as upper limits to the true values of sSFRs and we visualize this regime with the red vertical line of Fig. \ref{fig:sSFRIR1}. We find that our peak related to the passive population has a lower amplitude than the one found by \citet{Wetzel2011} when we impose our strict ${\rm SFRs = 10^{-1.5}}$ ${\rm M_{\odot} / {\rm yr}}$ cut \footnote{SDSS SED SFRFs are robust and in good agreement with other SFR indicators and cosmological simulations above the confidence limit of ${\rm SFRs = 10^{-1.5}}$ ${\rm M_{\odot} / {\rm yr}}$ \citep{Zhaoka2020}. Including objects below these limits would guarantee that the derived sSFRFs are incomplete.}. Furthermore, the peak associated to the quenched galaxies shifts slightly towards higher sSFRs with respect the parent sample which includes objects below the confidence limit. We demonstrate that the secondary peaks of the  ${\rm M_{\star}} = 10^{9.5} - 10^{10} \, {\rm M_{\odot}}$ and ${\rm M_{\star}} = 10^{10} - 10^{10.5} \, {\rm M_{\odot}}$ (top right panel of Fig. \ref{fig:sSFRIR1}) bins are found at $10^{-11}  \, {\rm yr^{-1}}$ and  $10^{-11.5}  \, {\rm yr^{-1}}$, respectively. These are well above the $10^{-12}  \, {\rm yr^{-1}}$ threshold value. In addition, the passive distribution of ${\rm M_{\star}} = 10^{10.5} - 10^{11} \, {\rm M_{\odot}}$ (bottom left panel of Fig. \ref{fig:sSFRIR1}) with ${\rm sSFRs} < 10^{-11.5} \, {\rm yr^{-1}}$ is enough to probe a non uni-modal distribution, without the need to involve any galaxies around ${\rm sSFRs} = 10^{-12.0} \, {\rm yr^{-1}}$. Thus, we suggest that the bi-modality found in SDSS is robust within the SFR/mass limits adopted in our work. However, at the last mass bin where AGN contamination is expected to be important and SDSS SFRs are more uncertain, the peak value (${\rm sSFRs} = 10^{-11.9} \, {\rm yr^{-1}}$) appears very close to the sSFR limit (${\rm sSFRs} = 10^{-12} \, {\rm yr^{-1}}$) and could be indeed an artefact of the shortcomings discussed above. We have to note that our sub-sample is not a representative of the overall/total galaxy distribution. It instead represents galaxies with ${\rm SFRs > 10^{-1.5}}$ ${\rm M_{\odot} / {\rm yr}}$ and  ${\rm M_{\star}} > 10^{9.0} \, {\rm M_{\odot}}$. \footnote{A higher stellar mass cut would result in a sSFR distribution that includes more quenched objects since the quenched fraction increases with increasing mass \citep{Fang2018}. The ${\rm M_{\star}} > 10^{9.0} \, {\rm M_{\odot}}$ limit, which represents the confidence mass limit of SDSS allows the selection of both quenched and star-forming objects. The mass bins we explore (e.g. ${\rm M_{\star}} = 10^{9.5} - 10^{10} \, {\rm M_{\odot}}$) involve objects above this stellar mass limit so the SFR cut is the one affecting the derived sSFRFs. Since we apply a SFR cut, some low sSFR galaxies are excluded from the analysis and this makes the amplitude of the secondary peak lower than the one found in the parent sample. In addition, the position of the passive distribution/peak moves towards higher sSFRs.}

For comparison we show the observed sSFR of \citet{Ilbert2015} from the combined COSMOS and GOODS surveys which includes solely {\it star-forming objects} selected by a main sequence definition using a color-color selection) represented by the orange solid line and orange dotted line (shifted by -0.25 dex\footnote{In addition to the original data of Ilbert et al. we plot the same results shifted by -0.25 dex in order to consider systematics between the different ${\rm SFR_{SED}}$, ${\rm SFR_{H \alpha}}$ and ${\rm SFR_{UV+IR}}$ indicators \citep{Ilbert2015,Dai2018,Caplar2019,Katsianis2020,Lower2020}.}) of Fig. \ref{fig:sSFRIR1}. We find that the star-forming peak given by \citet{Ilbert2015} is in good agreement with our results from SDSS within systematics between different SFR indicators \citep{Katsianis2017}.  \citet{Ilbert2015} suggested that the double-exponential and log-normal profiles provide good fits for their data with the double-exponential fit being more successful, especially at larger masses. The authors find that the shape of the sSFRF of objects highly depends on the mass, which we confirm in our work.

In conclusion, in Fig. \ref{fig:sSFRIR1} we show the bi-modal form of the SDSS sSFRF within the strict limits of $ {\rm SFRs > 10^{-1.5}}$ ${\rm M_{\odot} / {\rm yr}}$ and ${\rm M_{\star} > 10^{9}} \, {\rm M_{\odot}}$. The shape of the distribution of sSFRs has been supported by previous studies based on other data-sets \citep{Santini2009,Lenkic2016,dave2019}. In addition, we show how smoothly quenching is happening across different mass scales within these limits. Starting from the low-mass end, we demonstrate that most objects are star forming (being in excellent agreement with the double exponential form of the main sequence study of Ilbert et al.), but a small population that lies exactly at the limit of passive and star-forming objects arises  (${\rm sSFRs} \sim 10^{-11} \, {\rm yr^{-1}}$). Moving to intermediate-mass objects, a passive population of galaxies with low specific star formation rates (${\rm sSFRs} \sim 10^{-11.5} \, {\rm yr^{-1}}$) starts emerging and it is abundant enough to give a bi-modal shape to the sSFRF. At the highest mass bins, the quenched population surpasses the star forming, though the latter is still significant. In table \ref{tab_stepssfrf1} we report the results from our analysis. The question that arises is: Are State-of-the-art cosmological models qualitatively in agreement with the observed sSFRF within the ${\rm SFR/M_{\star}}$ limits we adopted?

\section{The sSFRF at different mass scales in cosmological models}
\label{simsbina}

\subsection{Cosmological simulations and Semi-analytic models}

In this work we explore a range of cosmological simulations. These assume various schemes, especially in terms of feedback prescriptions. The latter encloses the philosophy that is used from each group to quench galaxies and plays a major role in shaping the simulated objects, which otherwise would be extremely massive and star forming \citep{Dave2011,Katsianis2014}. We summarize the models considered below and in table \ref{tab:sim_runs}:

\begin{table*}
\centering
\resizebox{1.0\textwidth}{!}{%
\begin{tabular}{llccccccc}
  \\ \hline & Run & L & N$_{\rm TOT}$ & m$_{\rm DM}$ & sSFRs Shift  &
  Code & Feedback \\ & & [Mpc] &
  & [M$_{\rm \odot}$] & [dex] & & (Ref) \\ \hline
  & Illustris & 106.5 & $1820^3$ & 6.26$\times10^{6}$ &
  -0.3 &  AREPO & SNe: Winds, kinetic $+$ AGN: Radio mode-Mechanical, Quasar mode, Thermal/heating, radiation   \\ \hline
  & EAGLE & 100 &  $1504^3$ & 9.70$\times10^{6}$ & +0.2  & GADGET-3 & SNe: Thermal/heating  $+$ AGN: Thermal/heating  \\ \hline
  & Mufasa & 73.5 & $512^3$ & 9.70$\times10^{7}$ &
  -0.3 & GIZMO & SNe: Winds-Kinetic  $+$ AGN: Scheme to mimic Radio mode-heating \\ \hline
  & IllustrisTNG & 110.7 & $1820^3$ & 7.5$\times10^{6}$ &
  -0.2 & AREPO & SNe: Winds-kinetic $+$ AGN: Radio mode-Winds, Quasar mode-Thermal/heating \\ \hline
   & Simba & 148 & $1024^3$ & 9.70$\times10^{7}$ &
  0 to -0.3 & GIZMO & SNe: Winds-Kinetic  $+$ AGN: Radio mode-Xray heating, Quasar mode-Winds and kinetic \\ \hline \hline \hline
  & Shark & 309.73 & $1536^{3}$ & 2.21$\times10^{8}$ &
   - 0.3 & GADGET-2 & SNe: Winds-Kinetic  $+$ AGN: Radio mode-heating \\ \hline
  & ELUCID+L-Galaxies & 694.4 & $3072^{3}$ & 4.3$\times10^{8}$ &
   0 to + 0.1 & L-GADGET & SNe: Heating, winds $+$ AGN: Radio mode-Thermal/heating, Quasar mode-Winds  \\ \hline
\end{tabular}%
}
\caption{Summary of the different cosmological simulations used in this work. Column 1, run name; column 2, box size of the simulation in comoving Mpc; Column 3, total number of dark matter particles; Column 4 mass of the dark matter particle; Column 5, sSFR shift of the distribution needed to bring observed and simulated high star forming peaks to consonance; Column 6, code used; Column 7, combination of feedback implemented. }
\label{tab_stepssfrf1}
\end{table*}
 
\begin{itemize}

\item {\bf Illustris} \citep{Genel2014,Vogelsberger2014} consists a cosmological simulation run with the moving-mesh code AREPO \citep{springel2010}. Its sub-grid physics involve stochastic star formation, kinetic stellar feedback driven  by SNe explosions\footnote{Supernova energy is injected as kinetic energy to closeby SPH particles and subsequently decoupling the particles from hydrodynamic interactions to ensure their escape from the galaxy.}, super-massive black hole (SMBH) growth and related AGN feedback since cosmological simulations lack the resolution to resolve all these phenomena. The free parameters ($> 15$) are constrained by matching the results with the Cosmic Star Formation Rate density (CSFRD) evolution and the GSMF at $z \sim 0$ \citep{Vogelsberger2013}. SNe feedback is modelled in terms of galactic winds \citep{Lopez2020} which employ an energy-driven scheme \citep{PuchweinSpri12}. The  wind  particles are launched in the neighborhood of stellar particles \citep{Okamoto2010}  and are  kicked in a random directions, depositing their mass, momentum and thermal energy. In Illustris Black Holes (BHs) and AGN feedback rely on the implementations given by \citep{Springel2005,Sijacki2007,DiMatteo2008}. BHs are represented by collisionless, massive sink particles which grow in mass by accreting gas and merging with other BHs. Three different feedback processes emerge as a result of this growth including

\begin{itemize}

\item thermal back reaction: a fraction of the energy released by the accreted gas couples thermally to nearby gas,
\item mechanical feedback: AGN jets inflate hot, buoyantly rising bubbles in the surrounding halo atmosphere,
\item electro-magnetic/radiative back reaction: the photo-ionisation and photo-heating rates  of nearby plasma change due to the presence of the AGN radiation field.
\end{itemize}
At high BH accretion rates (quasar-mode) thermal feedback is the dominant mechanism, while for low accretion rates (radio-mode) the mechanical feedback is important \citep{Sijacki2008}. The radiative feedback is most effective for accretion rates close to the Eddington limit but is considered weak compared to both thermal and mechanical feedback contributions \citep{Vogelsberger2013}. In addition, \citet{Sijacki2008} using cosmological simulations have found that the SFRs of galaxies are not sensitive to the nature of the bubble feedback. 

\item {\bf EAGLE} \citep{Schaye2015}, run with an extended version of GADGET \citep{Springel2005}, adopts the stochastic thermal feedback scheme described in \citet{DVecchia2012}.  When a stellar particle reach the age of $3 \times 10^7 \, {\rm yr}$, it injects a spherical  thermal  input, increasing the internal energy of the neighboring  particles and giving them a temperature jump $\Delta T$. This makes the available cold gas less and reduces star formation. The fraction ($f_{th, max}$) of the energy output is typically absorbed mostly by galaxies with low metallicities and high densities \citep{Katsianis2017}. In addition to SNe feedback, EAGLE employs AGN feedback as well. Galaxies are seeded by BHs following \citet{Springel2005}, where seeds are placed at the center of every halo more massive than ${\rm M_{\star}} = 10^{10}$ M$_{\rm\odot}/h$ that does not already contain a BH. These grow by accretion of nearby gas particles or through mergers with other BHs. The gas accretion obeys the Bondi-Hoyle-Lyttleton formula following \citet{Rosas-Guevara2015} and assumes a radiative efficiency of $\epsilon_r = 0.1$. This will stochastically heat  neighboring  particles (i.e. thermal feedback). AGN feedback by construction in EAGLE quench star formation in massive galaxies and is responsible for reproducing the high mass end of the stellar mass function \citep{Furlong2014} and the super-massive black hole mass function \citep{Rosas2016}. Cosmological simulations often make a distinction between ‘quasar’- and ‘radio-mode’ BH feedback \citep{Croton2006,Sijacki2008}. The first occurs when the BH is accreting efficiently and comes in the form of a hot, nuclear wind, while the second operates when the accretion rate is low compared to the Eddington rate. EAGLE does not distinguish these two feedback modes unlike Illustris and its implementation is closer to a quasar-mode back reaction \citep{Schaye2015}. The rational behind this is to limit the number of feedback channels to the minimum required to match the observations. We note that \citet{Bower2017} showed that the AGN feedback in EAGLE keeps the surrounding gas heated. In this scheme a decreasing effectiveness of feedback from star formation leads to an increase in accretion into the black hole and thus black hole feedback will take over as star formation driven feedback fails to quench star formation. The above resembles a radio-mode back reaction instead of a quasar-mode behavior.

\item {\bf Mufasa} \citep{Dave2017} uses a modified  version  of  the code Gizmo \citep{Hopkins2015},  which relies on the GADGET-3 gravity  solver. Stellar feedback is modelled using outflows in the form of decoupled two-phase winds relying on the Feedback In Realistic Environments (FIRE) zoom in simulations \citep{Muratov2015}. Thus, kinetic  outflows  are ejected with energy and momentum input to the intergalactic medium. The model allows hydrodynamical interactions and cooling of all gas at all times, unlike Illustris. To  quench  massive  galaxies Mufasa follows the mass based quenching scheme given by   \citet{Gabor2012}, in which  a  halo  above a mass ${\rm M_{\rm quench}} =  (0.96 + 0.48z) \times 10^{12}$ ${\rm M_{\odot}}$, maintains  all  halo  gas at high temperature. This is intended to mimic the effects of the “radio mode” quenching that  counteracts  gas  cooling and prevents star formation \citep{Croton2006,Bower2017}. 

\item The {\bf IllustrisTNG} \citep{Pillepich2018} project is the successor of the Illustris simulations and includes an updated galaxy formation model with improvements to the SNe and AGN feedback \citep{Weinberger2017} prescriptions in order to address some shortcomings of the original Illustris run. The above implementations included enhanced feedback, especially for objects with $10^{12}-10^{14} \, {\rm M_{\odot}}$  halo  masses, Galactic Winds (GW) are injected this time isotropically, with larger wind, velocity and Energy Factors. So overall quenching is more efficient in IllustrisTNG than Illustris. This choice was driven by the fact that the last was unable to reproduce the $z < 1$ star formation rate density, high mass end of stellar mass function at $z \sim 0$ and color bi-modality. For  the case of the AGN feedback for high BH accretion rates relative to the Eddington  limit, it is assumed that a  fraction  of  the  accreted  rest  mass  energy  heats the  surrounding  gas  thermally.  For low accretion rates a pure kinetic feedback component is used that inputs momentum to the surrounding gas in a stochastic manner. A key difference with the original Illustris model is that the radio bubble feedback for low accretion rates is replaced by kinetic winds.

\item {\bf Simba} \citep{dave2019} is the successor of Mufasa. The employed metal-loaded winds rely on  \citet{Muratov2015} and \citet{Angeles2017a}. The most significant improvement is that instead of the mass  quenching feedback scheme \citep{Gabor2012}, BHs are included, seeded and grown. There are two channels for BH accretion: a Torque-limited component from cold gas and a Bondi component from hot gas. The resulting energy from the above process \citep{Angeles2017b} is used to drive feedback that quenches galaxies. The above is modeled via kinetic bi-polar outflows and X-ray heating. The kinetic feedback component kicks for high Eddington ratios in which AGN drive galactic winds with velocities of $ \sim 1000$ km/s \citep{Perna2017}. The X-ray heating back reaction is modeled by a jet mode in which AGN drive hot gas that heats the surrounding gas \citep{Fabian2012} for objects with low Eddington ratios. This way the observed dichotomy in BH growth is achieved \citep{Perna2017}. X-ray feedback is modeled by  heating from black holes following the scheme introduced by \citet{Choi2012}. According to \citet{dave2019} this heating has a minimal effect on the galaxy mass function, but provides an important factor to fully quench massive galaxies. In short, the AGN model employed in SIMBA has  some similarities  to  the  two-mode  thermal/kinetic  AGN  feedback  model  employed  in Illustris-TNG. However, there are some key differences like the fact that Illustris-TNG  uses a spherical  thermal  back-reaction  at  high Eddington ratios, while Mufasa employs a kinetic feedback.

\end{itemize}  

We have to note that cosmological simulations, like the ones discussed above, are computationally expensive since they involve both dark matter and gas elements/particles. This makes them restricted in terms of resolution (thus, are uncertain at the low mass end) and box-size (thus, have limited statistics for massive objects). Another limitation is that the parameters of the models that describe star formation and feedback depend highly on resolution \citep{Zhaoka2020} and usually there is no convergence between different resolutions \citep{Schaye2015}. Thus, it is debatable if any of these models are actually physical or they have the ``ideal'' combination of parameters {\it for the adopted resolution} in order to re-produce key observables like the GSMF. On the other hand, semi-analytic models (SAMs) use pre-calculated dark matter merger trees from N-body simulations and follow the formation of galaxies with simplified and observationally motivated, analytic prescriptions \citep{Hirschmann2012,Shark2019}. Since SAMs are less computationally expensive in the construction of galaxy samples, they can achieve much better statistics (by orders of magnitude) compared to those from cosmological hydrodynamical simulations. In our work we employ the following semi-analytic models:

\begin{itemize}  

\item The {\bf ELUCID} simulation \citep{Elucid2016} traces $3072^{3}$ dark matter particles in a periodic box of 644 Mpc. The N-body simulation was run with L-GADGET, a memory-optimized version of GADGET-2. It is constrained (in terms of initial conditions) by the re-constructed initial density field of SDSS DR 7 \citep{Elucid2014}. ELUCID has been used to study galaxy quenching \citep{Wang2018Elu}, galaxy intrinsic alignment \citep{Wei2018elu} and cosmic variance \citep{Chen2019elu} while it was also combined with an abundance matching method to evaluate galaxy formation models \citep{Yang2019}. In our work, we combine the merger trees taken from the ELUCID N-body simulations with the L-Galaxies Semi Analytic Model \citep[SAM, ][]{Luo16}. Galaxies are assumed to form at the centers of the dark matter haloes while the recipes that describe physical baryonic processes  (e.g.  star formation and metal production, SNe feedback, black hole growth and AGN feedback) are implemented by the SAM \citep{Fu2010,Fu2013}. The parameters are constrained by the observed galaxy stellar and HI (neutral hydrogen) and H2  (molecular hydrogen ) mass functions at $z = 0$. SNe feedback decreases the SFRs of the low mass objects by heating and ejecting the available gas within the halo \citep{Fu2013}, while the AGN model \citep{Croton2006} is extremely efficient in switching off cooling in massive haloes. Some advantages of this model is that it is resolution independent (with successful convergence results for satellites in massive haloes), while it includes environmental effects like cold gas stripping.

\item The {\bf Shark} semi-analytic model \citep{Shark2019} is merged with the  Synthetic UniveRses For Surveys (SURFS) simulation of $1536^{3}$ dark matter particles in a periodic box of 310 Mpc \citep{Elahi2018}. The N-body simulation was run with a memory-optimized version of GADGET-2 for a  $\Lambda$CDM  cosmology.  The reference simulation uses a Planck  cosmology \citep{Planck2016}  and  sample  scales  and  halo masses  down  to 1 kpc and ${\rm M_{\star}} = 10^{8} \, {\rm M_{\odot}}$, respectively.  The  haloes are tracked and identified with a state-of-the-art  6D  halo  finder  and  merger  tree  builder \citep{Elahi2019}.  Shark includes several  processes such as gas cooling \citep{croton06}, stellar feedback \citep{Lagos2013,Muratov2015} in terms of galactic winds/outflows that escape either the host galaxy or the halo,  radio-mode AGN feedback \citep{Croton2016} in the form of gas heating and star formation \citep{Krumholz2013}. The parameters of the model were constrained from the $z=0$, 1, 2 stellar mass functions, the z =  0  black  hole-bulge  mass  relation and the stellar mass vs size relation. \citet{Bravo2020} showed that Shark reproduces well the color bi-modality, though the transition from predominantly star-forming galaxies to passive objects happens at stellar masses that are larger than the ones found by the GAMA observations.

\end{itemize}

We have to note that in SAMs, unlike cosmological simulations, the dynamics of the baryonic component (gas and stars) and their interaction with dark matter are not followed directly. This is one of the main disadvantages of SAMs which use analytical prescriptions to model the above \citep{Benson2011,Hirschmann2012}. In SAMs we are forced to make strong assumptions about the geometry and dynamics of gas, while they are unable to probe details for the galaxy and halo structure. Hydrodynamic simulations can compensate for this limitation and investigate the resolved properties of galaxies (like the gas-phase metallicity, SFR and stellar mass distributions) within galaxies, at the expense of being computationally expensive. In our work we focus on the galaxy sSFRF (i.e. integrated SFRs and integrated stellar masses of galaxies) and thus SAMs, with their unparalleled statistics and large volumes can provide powerful tools to investigate the sSFR distribution. 

In conclusion, both SAMs and hydrodynamical simulations are forced to implement uncertain models that involve numerous parameters to approximate star formation or feedback from SNe and AGN, with SAMs having the advantage of better statistics in larger volumes, while simulations having the advantage that they follow gas dynamics and do not need to employ analytic prescriptions to do so. There are differences and similarities between different collaborations. Some concordances  of great importance are that they tuned the parameters of their models to the observed $z \sim 0$ GSMF while SNe feedback decreases SFRs in low-mass objects and AGN is tuned to quench galaxies at the high-mass end.

\begin{figure*}
    \,
  \,
  \centering
  \vspace{0.16cm}
  \includegraphics[scale=0.97]{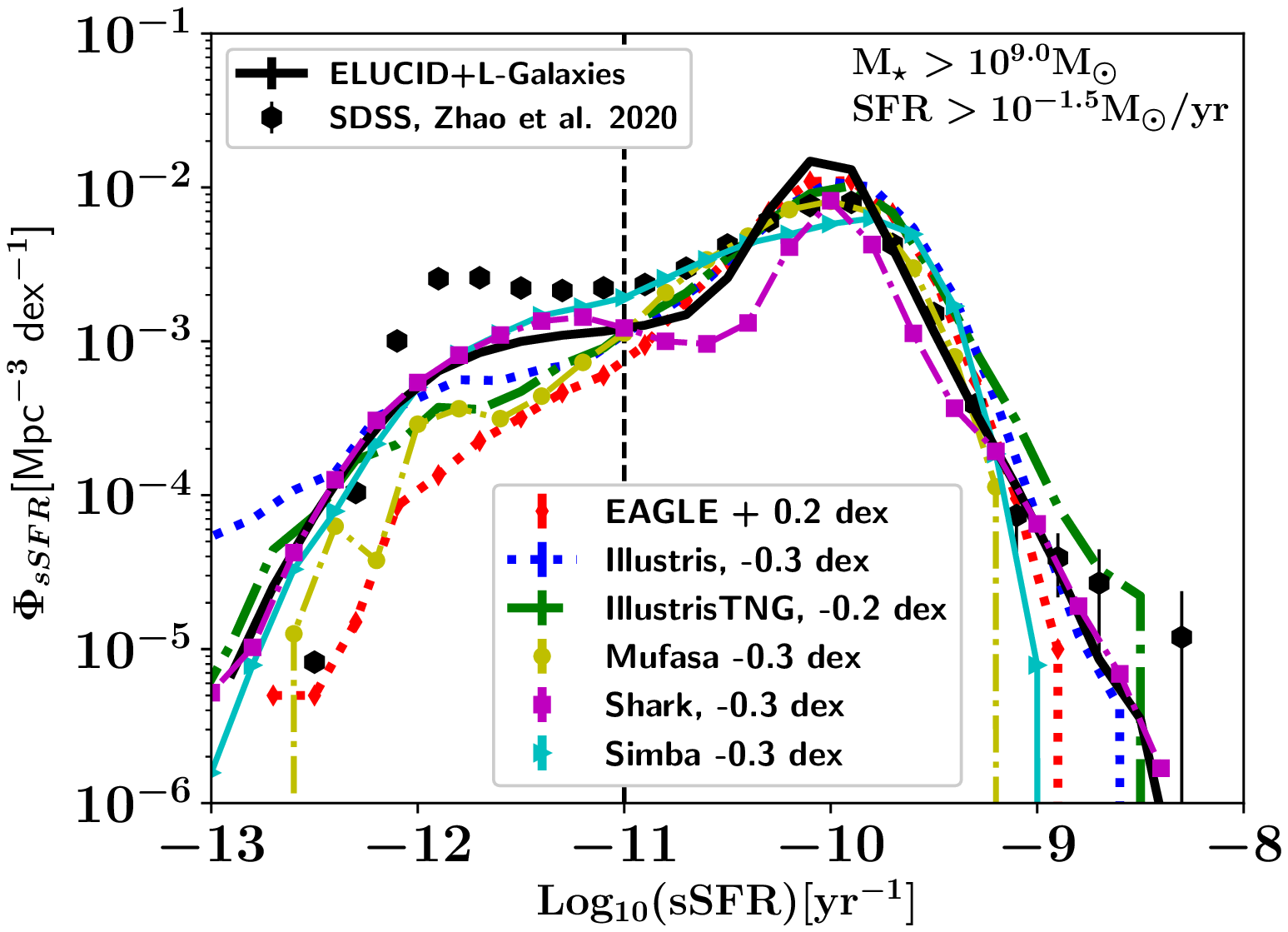}
  \vspace{-1.45cm}
\caption{ The sSFR in cosmological simulations: Illustris \citep[blue dotted line, ][]{Genel2014}, EAGLE \citep[red diamond line, ][]{Schaye2015}, Mufasa \citep[yellow line with circles, ][]{Dave2017} , IllustrisTNG \citep[green dashed-dotted line, ][]{Pillepich2018}, Shark \citep[magenta line with squares, ][]{Shark2019}, Simba \citep[cyan line with triangles,][]{dave2019} and ELUCID+L-Galaxies \citep[black solid line,][]{Luo16}. Black hexagons represent the observed (Eddington bias corrected) sSFRF from \citet{Zhaoka2020}. We see that the bi-modality found in observations is not reproduced for a range of cosmological models (Illustris, Mufasa, EAGLE, IllustrisTNG), with the ELUCID+L-Galaxies (described by the black solid line), Shark (magenta line with squares), Simba (cyan line with triangles) performing better qualitatively, since a bi-modality of sSFRs is emerging. However, even the latter have quantitatively lower number density peaks for quenched objects.}
\label{BinnedSFRFtotal}
\end{figure*}

\begin{figure*}
\centering
\includegraphics[scale=0.47]{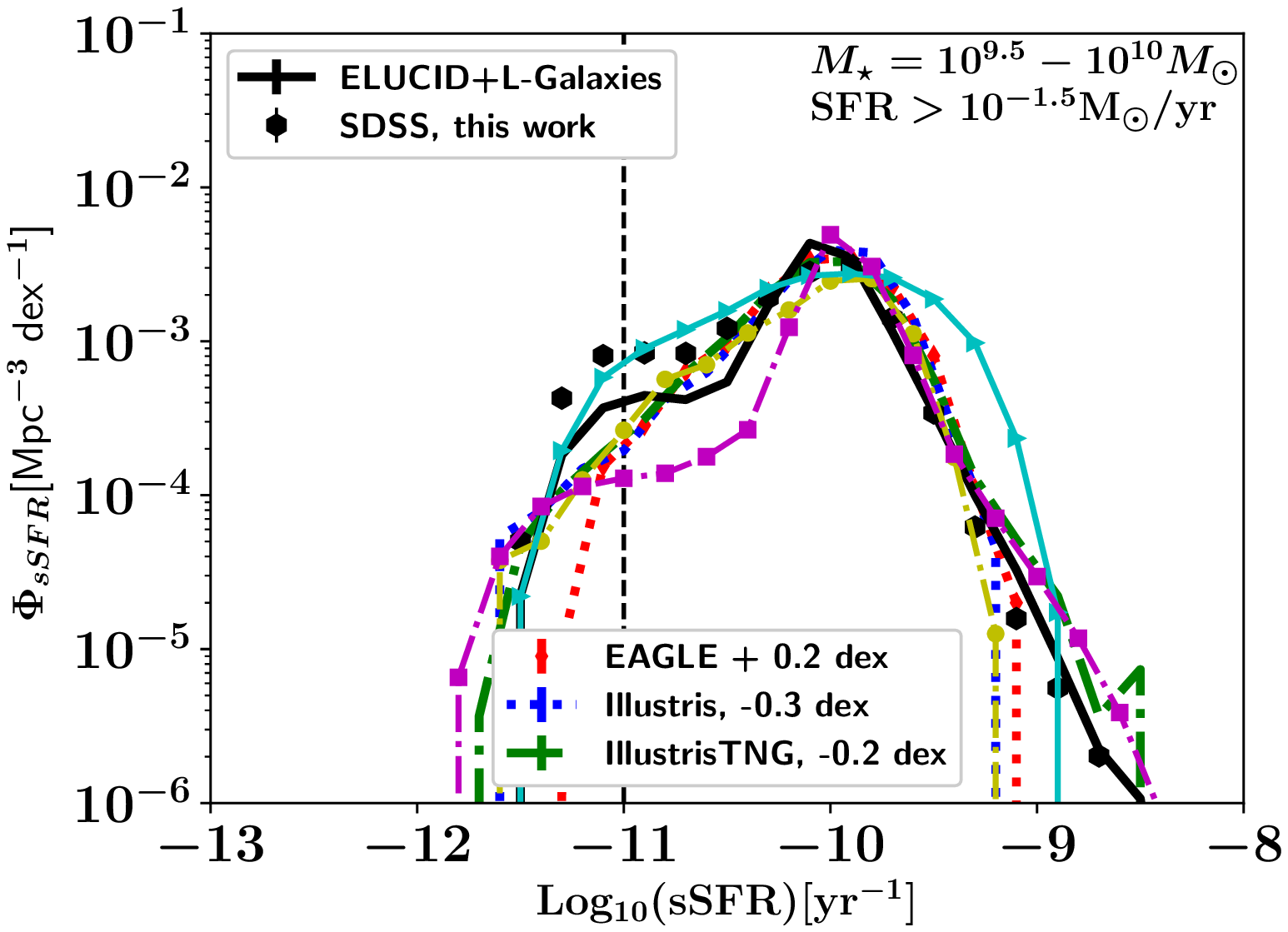}
\includegraphics[scale=0.47]{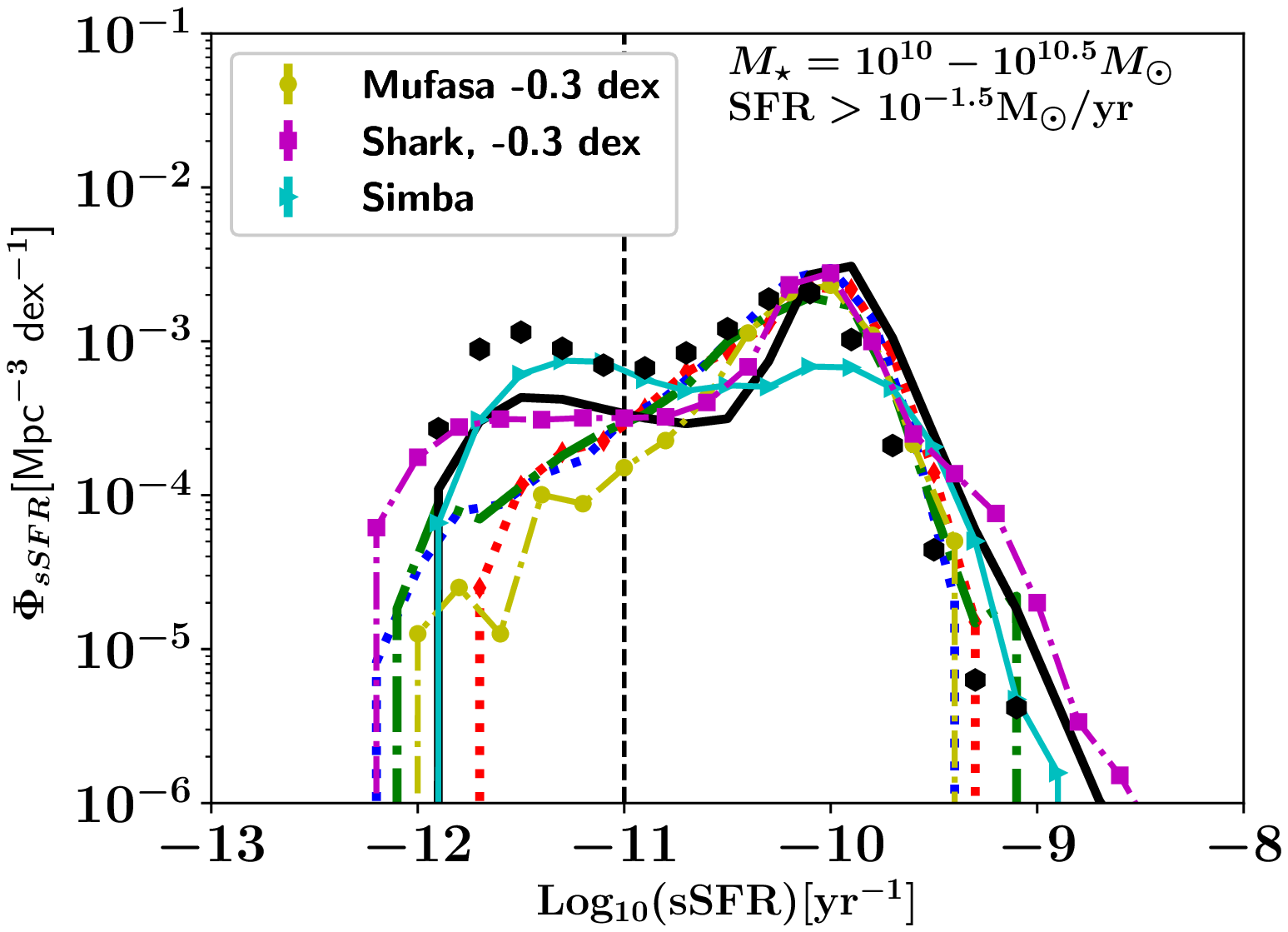} \\
\vspace{-0.76cm}
\includegraphics[scale=0.47]{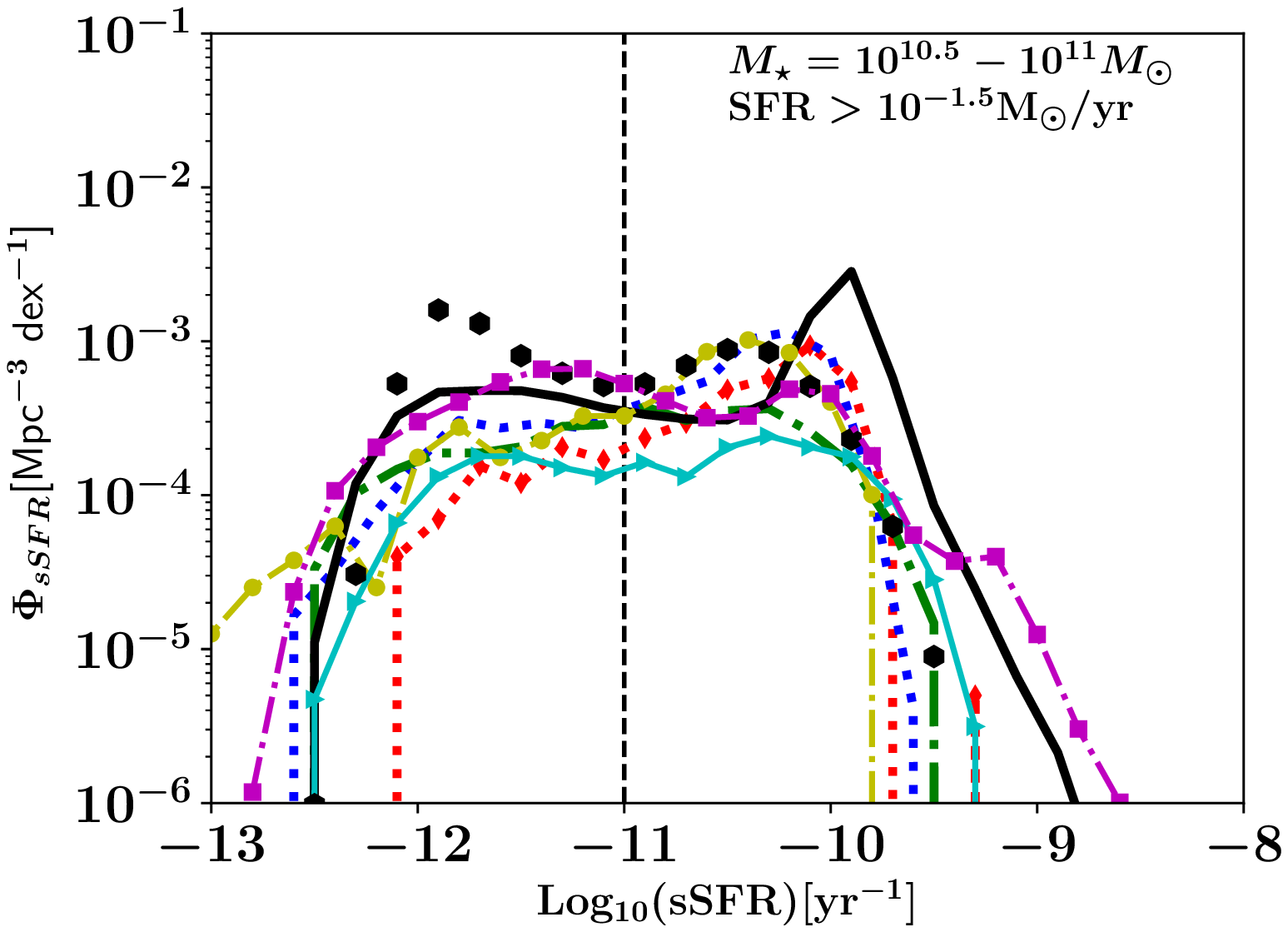}
\includegraphics[scale=0.47]{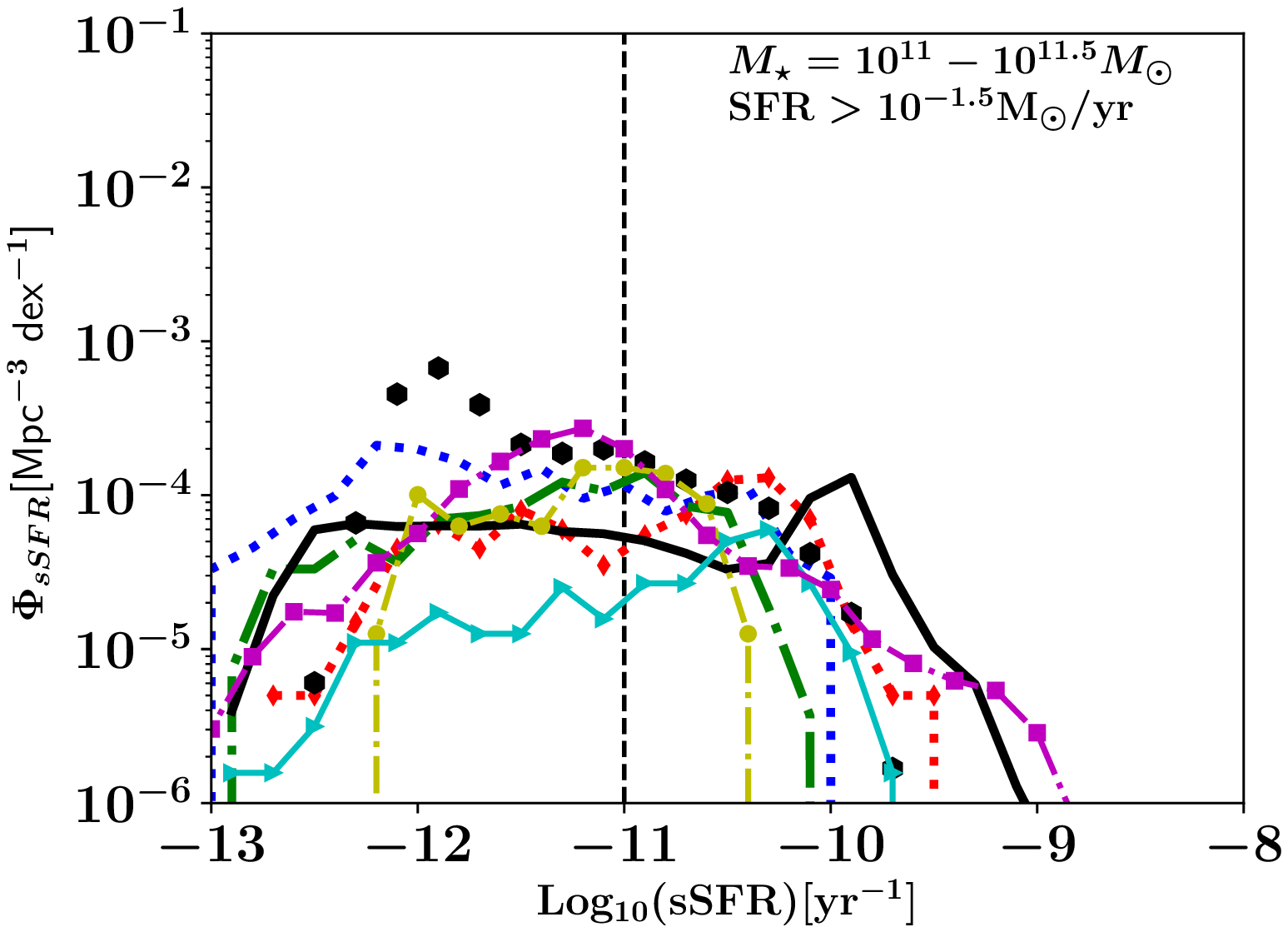}
\vspace{-0.45cm}
\caption{The sSFR at the mass scales/mass intervals of $10^{9.5} - 10^{10.0} \, {\rm M_{\odot}}$ (top left panel), $10^{10.0} - 10^{10.5} \, {\rm M_{\odot}}$ (top right panel), $10^{10.5} - 10^{11.0} \, {\rm M_{\odot}}$ (bottom left panel) and $10^{11.0} - 10^{11.5} \, {\rm M_{\odot}}$ (bottom right panel) for $ {\rm SFRs > 10^{-1.5}}$ ${\rm M_{\odot} / {\rm yr}}$ galaxies in cosmological simulations: Illustris \citep[blue dotted line][]{Genel2014}, EAGLE \citep[red line with diamonds][]{Schaye2015}, Mufasa \citep[yellow line with circles][]{Dave2017} , IllustrisTNG \citep[green dashed-dotted line][]{Pillepich2018}, Simba \citep[green line with triangles][]{dave2019}, Shark \citep[magenta line with squares, ][]{Shark2019} and ELUCID+L-Galaxies \citep[black solid line][]{Luo16}. Black hexagons represent the Eddington bias corrected sSFRF from SDSS at different mass bins present at Fig. \ref{fig:sSFRIR1}.}
\label{fig:sSFRIR11}
\end{figure*}

\subsection{The Simulated versus the Observed sSFRF}
\label{ObsvsSims}

We note that the comparisons we perform at this work are strictly done within the same limits for both observations and simulations (${\rm M_{\star}} > 10^{9} $ ${\rm M_{\odot}}$ in terms of stellar mass and SFR $> 10^{-1.5}$ ${\rm M_{\odot} / {\rm yr}}$). We also stress that the simulations discussed typically produce SFRFs \citep{Dave2017,Katsianis2017,Zhaoka2020}  and GSMFs (by construction) in excellent agreement with observations and are supposed to not be affected by resolution effects within the above SFR and stellar mass limits.

\begin{figure*}
    \,
  \,
  \centering
  \includegraphics[scale=0.85]{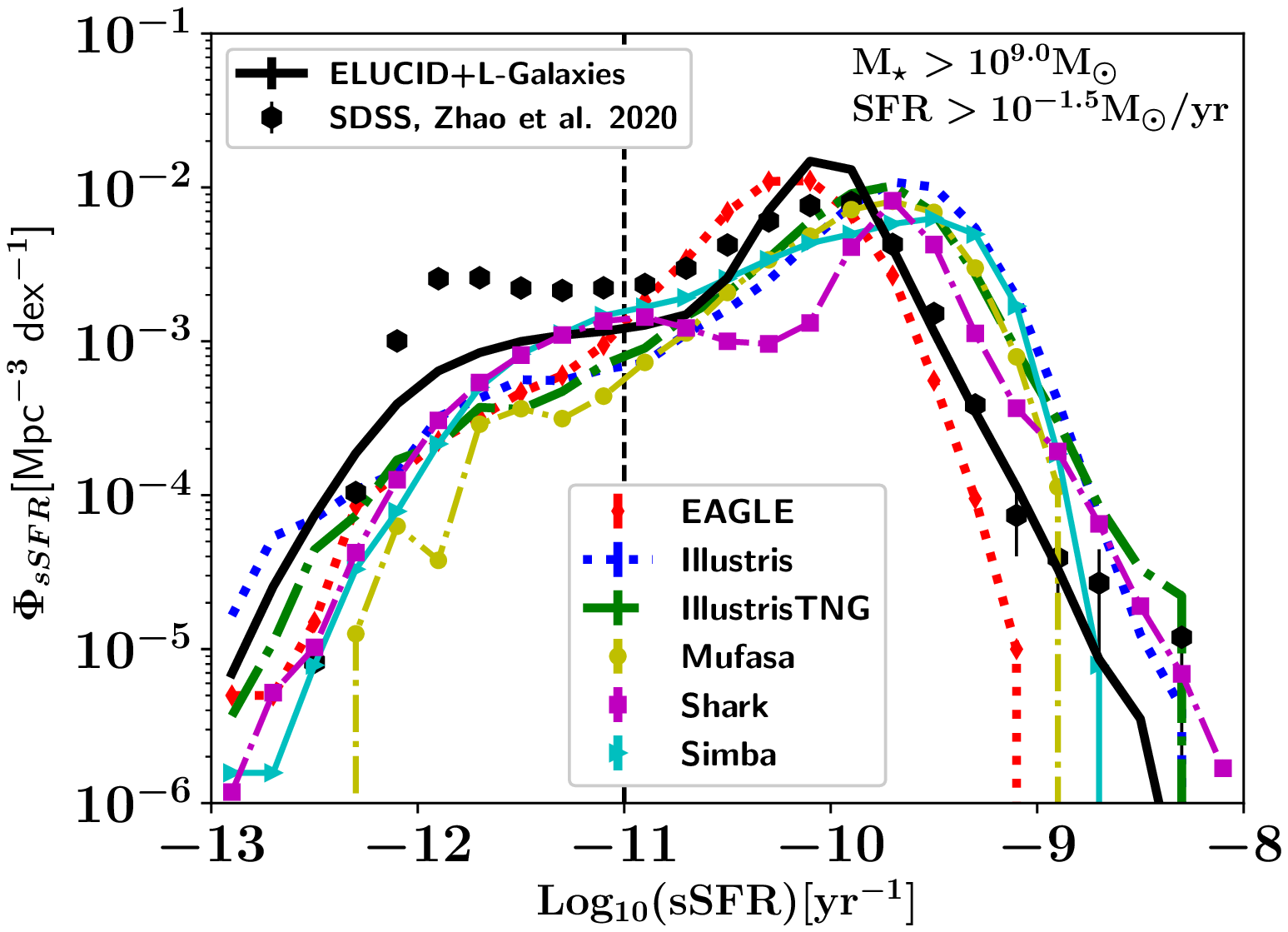}
  \vspace{-0.35cm}
  \includegraphics[scale=0.36]{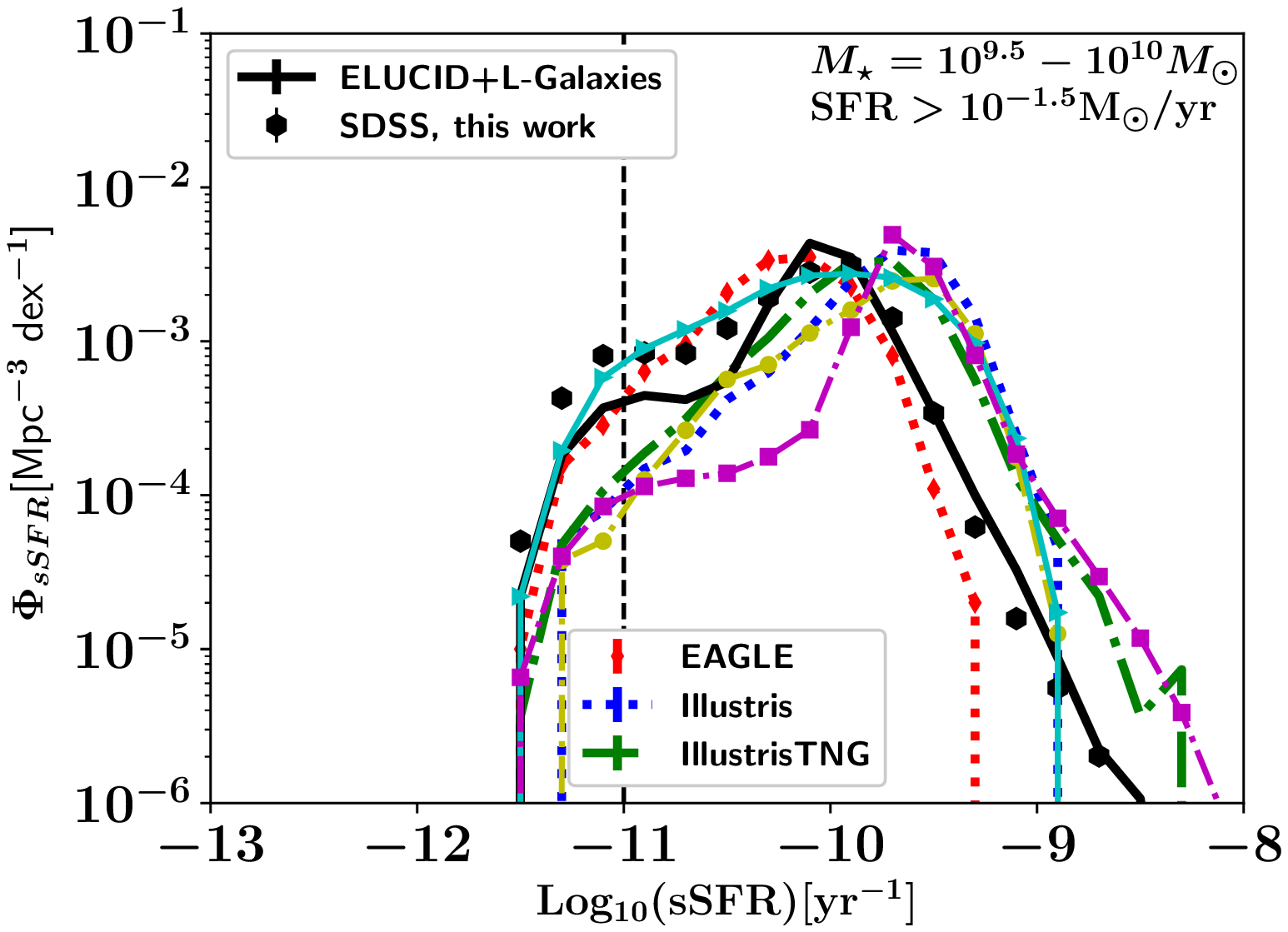}
\includegraphics[scale=0.36]{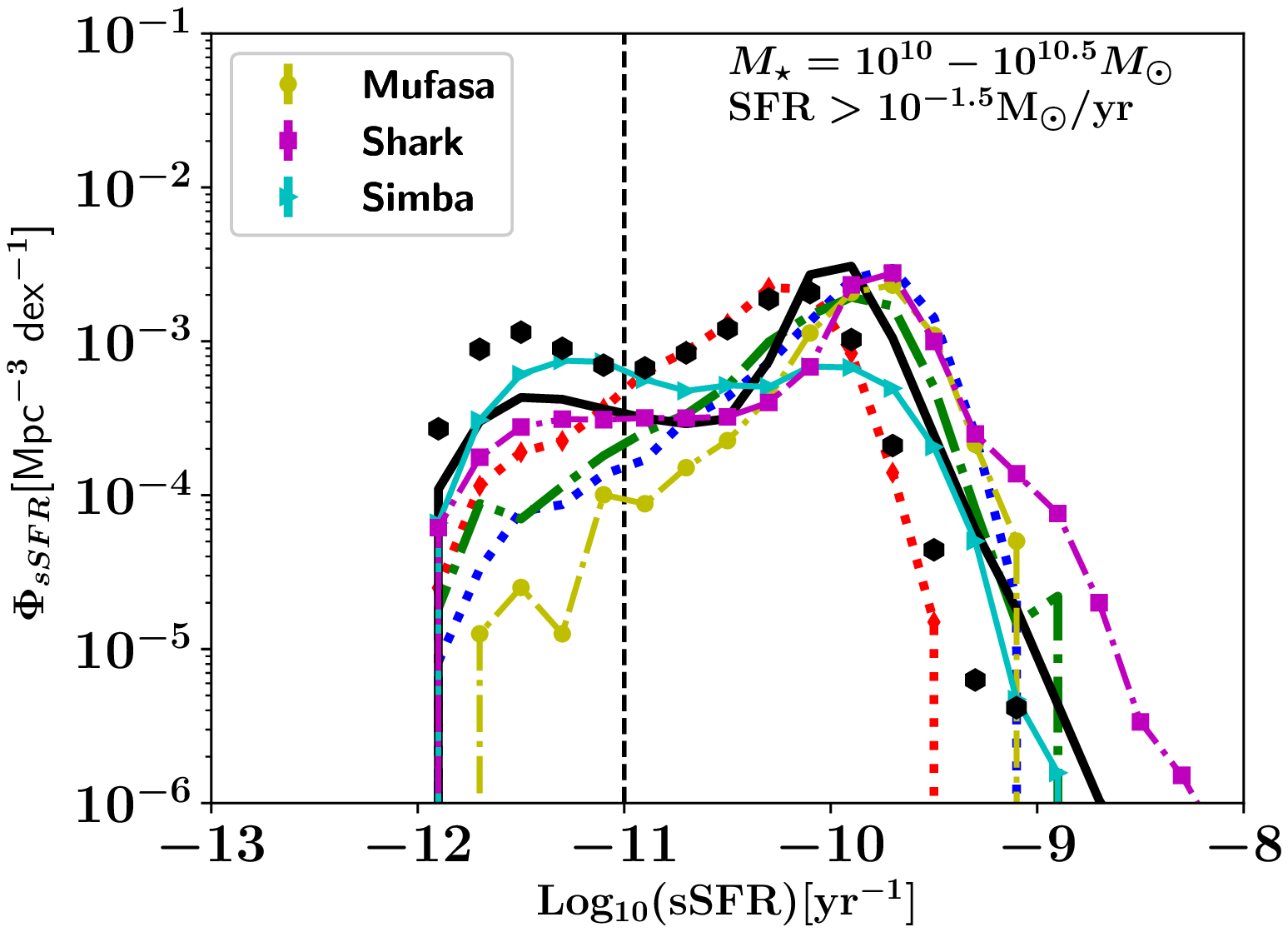}
\includegraphics[scale=0.36]{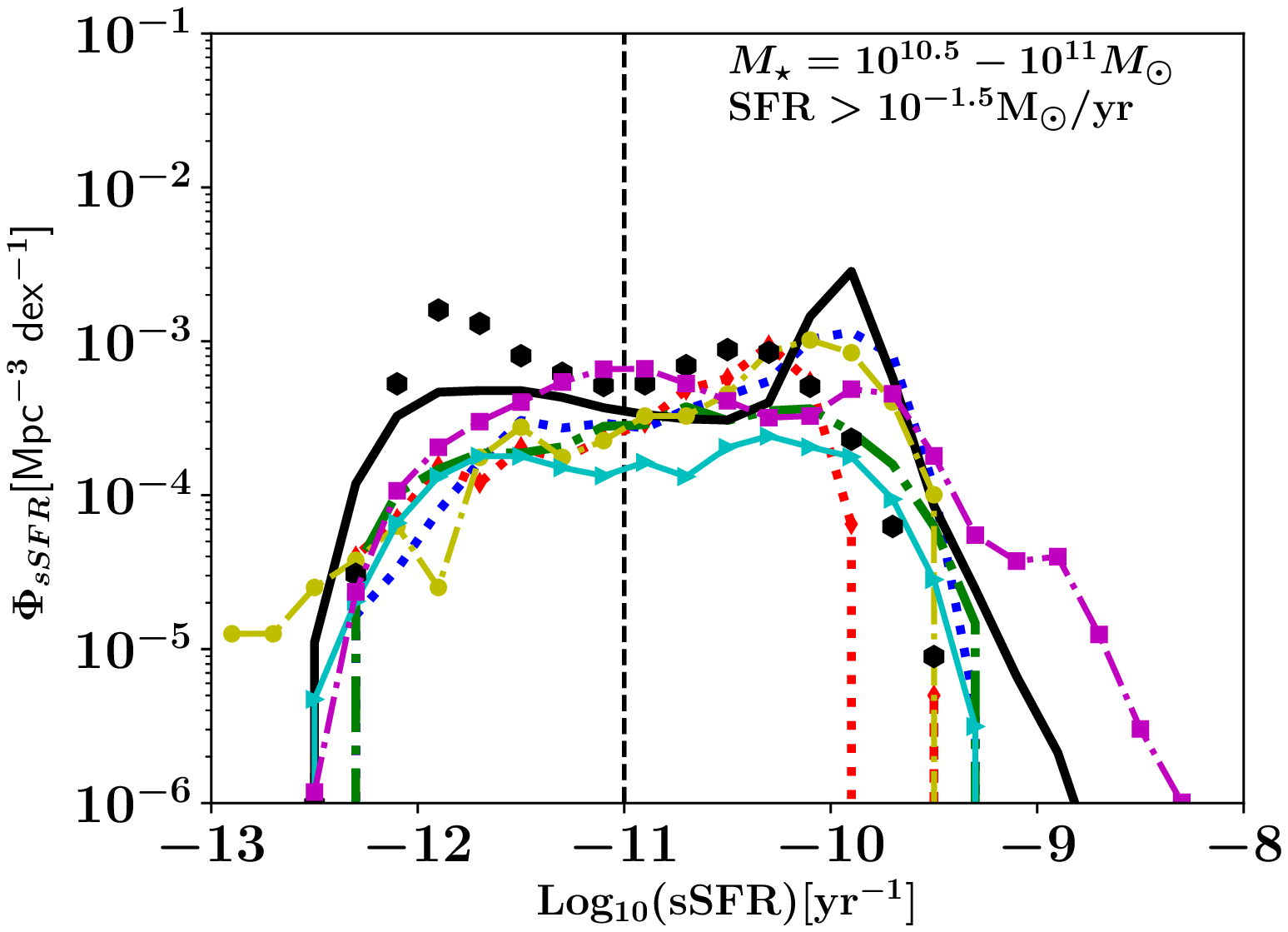}
  \caption{Same as Fig. \ref{BinnedSFRFtotal} and Fig. \ref{fig:sSFRIR11} but no re-scaling was employed for the models.}
\label{BinnedSFRFtotalno}
\end{figure*}

Before comparing the different simulations with the observations present at Fig. \ref{fig:sSFRIR1} we investigate in Fig. \ref{BinnedSFRFtotal} if models produce the bi-modal sSFRF described in \citet{Zhaoka2020} (black hexagons). Illustris is represented by the blue dotted line, EAGLE by the red line with diamonds, Mufasa is described by the yellow line with circles, IllustrisTNG by the green dashed-dotted line, Shark by the magenta line with squares, Simba by the green line with triangles and ELUCID+L-Galaxies by the dark solid line. In order to be able to perform the comparison qualitatively (the shape of the sSFRF) and be able to visually follow the lines of the plot we shift the originals lines in the x-axis by -0.3 dex, +0.2 dex, -0.3 dex, -0.3 dex, -0.3 dex, -0.3 dex and +0.0 dex, respectively\footnote{Here, we re-scale the whole sSFRF which is built using the remaining simulated/intrinsic SFRs and ${\rm M_{\star}}$ after applying our confidence limit cuts (no re-scaling was applied to the intrinsic properties). We explored how the shape changes (not shown in this work) when the sSFRFs are built using the re-scaled sSFRs (cuts were applied to the re-scaled properties this time) and find minimal differences between the resulting distribution.}. We note that any re-scaling is within the systematic uncertainties of different SFR indicators at low redshifts \citep{Katsianis2015,Davies2016,Lower2020} and it is a common practice used in order to facilitate qualitative comparisons between different data-sets and simulations \citep{McAlpine2017,Katsianis2019,Donnari2019}. For example, EAGLE SFRs are reported to be 0.2 dex lower than observations \citep{Katsianis2017,McAlpine2017}.  Following, \citet{McAlpine2017} and \citet{Katsianis2019}  we shift the simulated sSFRFs by +0.2 dex to focus on the shape of the distribution and facilitate a qualitative comparison with SDSS and the other cosmological simulations considered. On the other hand, Mufasa \citep{Dave2017} has larger SFRs with respect to the H$\alpha$ observations of \citet{Gunaw2013} by 0.3 dex, so we shift the sSFRFs  in Mufasa by this value (-0.3 dex). We perform a quantitative comparison without any re-scalings in Fig. \ref{BinnedSFRFtotalno}.

Starting from Fig. \ref{fig:sSFRIR11} and the qualitative comparison we see that Illustris, EAGLE, Mufasa and IllustrisTNG resemble a uni-modal double-exponential Gausian \citep{Ilbert2015,Katsianis2019} and while they are quite successful at reproducing the star-forming population and the related peak, they are unable to reproduce the secondary peak of quenched objects. The ELUCID+L-Galaxies model (described by the black solid line), Shark (magenta line with squares) and the Simba simulation (cyan line with triangles) perform better qualitatively with respect to the observations since a modest presence of a passive population emerges. However, even the latter have quantitatively lower number density of quenched objects compared to the SDSS observations.

In Fig. \ref{fig:sSFRIR11} we present the sSFRF at different stellar mass intervals of ${\rm M_{\star}} = 10^{9.5} - 10^{10.0} \, {\rm M_{\odot}}$ (top left panel), ${\rm M_{\star}} = 10^{10.0} - 10^{10.5} \, {\rm M_{\odot}}$ (top right panel), ${\rm M_{\star}} = 10^{10.5} - 10^{11.0} \, {\rm M_{\odot}}$ (bottom left panel) and ${\rm M_{\star}} = 10^{11.0} - 10^{11.5} \, {\rm M_{\odot}}$ (bottom right panel) of the cosmological simulations compared to our observational constraints. We see that Illustris (shifted by -0.3 dex), EAGLE (+0.2 dex), IllustrisTNG (-0.2 dex) and Mufasa (-0.3 dex) are in excellent qualitative agreement with each other at the ${\rm M_{\star}} = 10^{9.0} - 10^{9.5} \, {\rm M_{\odot}}$,  ${\rm M_{\star}} = 10^{9.5} - 10^{10} \, {\rm M_{\odot}}$ and  ${\rm M_{\star}} = 10^{10} - 10^{10.5} \, {\rm M_{\odot}}$ stellar mass bins. The above models reproduce successfully the star-forming peak of our SDSS observations. However, besides the differences between the different collaborations in quenching methods all the above models are unable to produce an adequate passive population within the strict limits we adopt and are unable to generate its associated peak. On the other hand, Simba (cyan line with triangles, original shifted by +0 dex), Shark (Magenta line with squares, -0.3 dex)  and L-Galaxies  (black solid line, + 0 dex) are able to reproduce bi-modal distributions at most stellar mass bins.

\citet{Katsianis2019} performed comparisons between the sSFRFs of \citet{Ilbert2015} and EAGLE. As discussed in section \ref{SFRSSDSS} the study of Ilbert et al included {\it star-forming objects only} and excluded quenched galaxies by using a color-color selection. An agreement was demonstrated  between simulations and observations without the need to exclude any ``passive'' objects from the simulation. Since there was not any necessity to apply a separation in order to achieve this agreement, it was implied  that EAGLE does not produce many quenched galaxies within the sSFR $=$ $10^{-11}$ to $10^{-13}$ ${\rm yr^{-1}}$ range. The scarcity of passive objects within these regimes for the reference simulation of EAGLE has been noted in other studies  as well \citep{Katsianis2020,Trvcka2020,Baes2020}.  \citet{Furlong2014} and \citet{Trayford2017} pointed out that a large number of the simulated galaxies in EAGLE (especially at the low-mass end) are influenced by numerical effects and poor particle sampling. This effect is responsible for low star formation rates for these un-resolved objects which appear quiescent with SFRs = 0. Their presence increases the passive fraction. However, this limitation is not enough to raise (as an artefact) an adequate passive population and a bi-modality for the simulated sSFRF, similar to the one found in SDSS (Fig. \ref{tab:sim_runs} and Fig. \ref{fig:sSFRIR11}). We find that 20 $\%$ of simulated galaxies at ${\rm M_{\star}} = 10^{9.5} - 10^{10} \, {\rm M_{\odot}}$, 18 $\%$ at the ${\rm M_{\star}} = 10^{10} - 10^{10.5} \, {\rm M_{\odot}}$, 17 $\%$ at the ${\rm M_{\star}} = 10^{10.5} - 10^{11} \, {\rm M_{\odot}}$ and 13 $\%$ at ${\rm M_{\star}} = 10^{11} - 10^{11.5} \, {\rm M_{\odot}}$ have SFRs equal to 0. It is a common practice \citep{dave2019} for these objects, since they are considered a by-product of limitations of resolution, to be granted artificially higher sSFRs and to be added at the last bin considered by the analysis (in our case sSFR $=$ $10^{-12.9} {\rm yr^{-1}}$). We stress that this is not enough to display a similar sSFR distribution as seen in SDSS, since the other sSFR bins representing passive objects (e.g. sSFR = $10^{-11}-10^{-12.7} {\rm yr^{-1}}$)  would remain unaffected by the above procedure (i.e. we would produce only an artificial peak and not the total secondary gaussian distribution). Following another approach we could grant an artificial gaussian distribution of SFRs to the objects with SFR = 0, {\it below} the SFR resolution limit of the simulation \citep[e.g. SFRs within $10^{-4}-10^{-5}$ ${\rm M_{\odot} / {\rm yr}}$, Fig. 8][]{Donnari2019}. However, these objects with artificially granted SFRs would be excluded from our analysis which adopts a strict limit ($ {\rm SFRs > 10^{-1.5}}$ ${\rm M_{\odot} / {\rm yr}}$) both in SDSS and simulations (i.e. un-resolved objects with SFRs = 0 would be granted SFR values below the resolution limit and thus would not affect the comparisons performed in Fig. \ref{BinnedSFRFtotal} and Fig. \ref{fig:sSFRIR11}.). We note that EAGLE represents the model, within our study, with the least active objects (distribution appears shifted towards lower sSFRs with respect the observations and the rest simulations). The peak representing the main sequence appears quantitatively 0.2 dex lower (at sSFR = $10^{-10.2} {\rm yr^{-1}}$) than the one found in SDSS observations (at sSFR = $10^{-10.0} {\rm yr^{-1}}$) and 0.5 dex lower than the one found in Illustris  (Fig. \ref{BinnedSFRFtotalno}), showing that the thermal feedback scheme employed in EAGLE is very effective. We note that the AGN feedback scheme, which does not separate between radio mode and quasar mode (subsection \ref{simsbina}), does not yield results qualitatively different from other models like Illustris-TNG (Fig. \ref{BinnedSFRFtotal} and Fig. \ref{fig:sSFRIR11}) that uses a more sophisticated prescription.

Similarly to \citet{Katsianis2019}, \citet{Dave2017} demonstrated that Mufasa is successful at reproducing the observations of \citet{Ilbert2015} and suggested that this implies that the model is able to quench galaxies successfully. As suggested in the previous paragraph, we believe that this comparison does not provide evidence of a successful quenching scheme since these observations do not include passive objects by construction. The comparison is strictly performed for galaxies of sSFR $=$ $10^{-11}$-$10^{-9}$ ${\rm yr^{-1}}$ (i.e. star-forming objects). The authors pointed that there is still a notable number of star-forming galaxies even at the highest stellar masses in Mufasa. Our analysis agrees quite well with this statement and supports the idea of having numerous star-forming galaxies even at the highest mass regime, with quenched galaxies however being more dominant (bottom panels of Fig. \ref{fig:sSFRIR11}). We note that the quenching mechanism adopted in Mufasa combined with limitations in resolution are responsible for totally halting star formation (SFR = 0) for 40 $\%$ of the simulated objects with $M_{\star} > 10^{9} \, {\rm M_{\odot}}$. These ``dead'' objects if allowed to retain some level of star formation (advances in resolution and especially modeling are required for this step) can potentially generate a secondary gaussian distribution similar to SDSS. As discussed in the previous subsection, cosmological models that attempt to follow both dark matter and gas suffer from limitations in resolution and box-size. We have to note that Mufasa is the simulation, within our work, with the lowest resolution and smallest volume. The numbers of objects at the high- and low- mass ends can be affected by these shortcomings and this in return can impact the retrieved sSFRF. If high mass and passive objects are under-represented in a simulation then the sSFR distribution describing the quenched population will unavoidably be incomplete. Besides these limitations and the fact that Mufasa does not employ a full treatment for the AGN feedback (i.e. uses just a scheme to resemble a radio-mode back reaction via heating) it produces qualitatively (Fig. \ref{BinnedSFRFtotal} and Fig. \ref{fig:sSFRIR11}) and quantitatively (Fig. \ref{BinnedSFRFtotalno}) sSFRFs in perfect agreement with Illustris which has 16 times higher mass resolution and 3 times larger volume. This comparison is a good example of how two completely different cosmological simulations with different feedback schemes, resolutions and volumes produce similar results due to a thoughtful tuning of their subgrid Physics for their adopted resolutions.

Contrary to Mufasa (yellow line with circles), its successor Simba (cyan line with triangles) produces a different distribution qualitatively and is able to produce a clear bi-modality for the ${\rm M_{\star}} = 10^{10} - 10^{10.5} \, {\rm M_{\odot}}$ and ${\rm M_{\star}} = 10^{10.5} - 10^{11} \, {\rm M_{\odot}}$ stellar mass bins. We consider this a success, possibly related to the updated model and the inclusion of a quenching feedback prescription via AGN. In  addition to the kinetic  AGN  back reaction, Simba also includes an X-ray feedback input. The importance  of  the X-ray heating  has  been  explored  in zoom  simulations  by \citet{Choi2012},  showing  that  it can  potentially  drive  the  quenching  of  massive  galaxies. Simba is the first cosmological-volume simulation to include such a mechanism and according to \citet{dave2019}, this heating besides the fact that it has a minimum effect on the GSMF, it provides an important factor to fully quench galaxies. Without the X-ray feedback component and only the kinetic back reaction from jets, Simba would not generate an adequate number of objects below sSFR $=$ $10^{-11}$ ${\rm yr^{-1}}$, and instead would generate mostly galaxies between sSFR $=$ $10^{-10}$ ${\rm yr^{-1}}$ and sSFR $=$ $10^{-9}$ ${\rm yr^{-1}}$ (like Illustris, or EAGLE). So the X-ray feedback plays a major role in Simba in reproducing quenched galaxies. \citet{dave2019} demonstrated that a bi-modal form emerges for the histograms of the simulated sSFRs and this is in agreement with the observed distributions obtained from the GALEX-SDSS-WISE  Legacy  CatalogS \citep[GSWLC, ][]{Salim2016}. The difference between Mufasa and Simba is that due to the updated AGN feedback mechanism, the Simba simulation has less star-forming objects, especially at higher masses. The affected galaxies have ``moved'' towards lower sSFRs and a bi-modality has emerged. However, this approach decreased significantly the star-forming population and its associated peak has a lower number density than the one found in our observations. We stress that in Simba 40 $\%$ of galaxies at ${\rm M_{\star}} = 10^{10.5} - 10^{11} \, {\rm M_{\odot}}$ and 60 $\%$ at ${\rm M_{\star}} = 10^{11} - 10^{11.5} \, {\rm M_{\odot}}$ have SFRs equal to 0. The above demonstrates the high efficiency of the kinetic and X-ray feedback prescriptions adopted in Simba for high mass objects\footnote{In contrast, the total fraction of ${\rm M_{\star}} > 10^{9} \, {\rm M_{\odot}}$ galaxies in Simba with SFR = 0 is only 18 $\%$.}. We also see the importance of the quenching mechanisms in totally pausing SF in simulated galaxies and generating ``dead'' objects with SFR = 0. A more moderate quenching mechanism,  would allow more objects to retain a low level of non-zero SFR, a larger normalization for the sSFRF and better quantitative agreement with observations. 

%On the other hand, Mufasa which has the same resolution as Simba but a different approach in quenching galaxies has a fraction of galaxies with SFRs = 0 of 39 $\%$ at ${\rm M_{\star}} = 10^{10.5} - 10^{11} \, {\rm M_{\odot}}$ and 52 $\%$ at ${\rm M_{\star}} = 10^{11} - 10^{11.5} \, {\rm M_{\odot}}$. The different percentages of objects with zero SFR seen in Simba with respect Mufasa at the high mass end has its origin to the different feedback prescriptions between the two models. 

Furthermore, we demonstrate that IllustrisTNG sSFRFs (dashed green lines of Fig. \ref{fig:sSFRIR11})  are not bi-modal at any mass bins considered. Illustris performs similarly to TNG except in the last stellar mass bin (bottom panel of Fig \ref{fig:sSFRIR11}). TNG, while keeping the good resolution and statistics of Illustris, includes a range of improvements with respect to the original illustris which made the updated model much more successful in terms of replicating key observables. One of those is the color bi-modality of SDSS \citep{Nelsoncolor}. As we suggested in the introduction, color and sSFRs are directly related so it would be expected that this success would be imprinted in the TNG sSFRFs as well. However, we obtain a uni-modal distribution for the simulation \footnote{\citet{Donnari2019} argued that a color bi-modality can co-exist with a uni-modal sSFR or SFR distribution. It is indeed debatable if galaxy colors and sSFRs have totally a one on one relation. We note that the un-resolved objects with SFRs equal to 0 are considered to be most of the red sequence in the TNG100 model.}. We note that the comparison between SDSS and TNG performed at Fig. \ref {fig:sSFRIR11} is strictly done well above the SFR resolution limits of the simulation ($ {\rm SFRs \sim 10^{-3}}$ ${\rm M_{\odot} / {\rm yr}}$) and thus even by granting artificial SFR greater than 0 (but lower than the resolution limit) to the numerous ``dead'' un-resolved objects, still these SFRs would be well below the $ {\rm SFRs \sim 10^{-1.5}}$ ${\rm M_{\odot} / {\rm yr}}$ threshold we adopted in our study and would result to ${\rm sSFRs} < 10^{-13} $  ${\rm yr^{-1}}$ \citep[e.g.][granted these objects SFRs within $10^{-4}-10^{-5}$ ${\rm M_{\odot} / {\rm yr}}$]{Donnari2019}.  This points to the direction that the discrepancy seen between the SDSS observations and TNG has its roots in modeling and not limits in resolution.

Finally, we show that the ELUCID+LGalaxies (black solid line of Fig. \ref{fig:sSFRIR11}) and Shark (magenta line with squares of Fig. \ref{fig:sSFRIR11}) models show promising bi-modal sSFRFs. The success of L-Galaxies is due to the fact that it allows passive objects to retain a low level of star-forming activity and generate a secondary distribution of passive objects and the associated peak. We note that the total number of galaxies with SFR = 0 is only 2 $\%$, thus numerous objects with low SFRs emerge instead of ``dead'' galaxies and a bi-modality arises. However, ELUCID+LGalaxies has typically sharper peaks found than observations, while the Shark passive population in the ${\rm M_{\star}} = 10^{10} - 10^{10.5} \, {\rm M_{\odot}}$ and ${\rm M_{\star}} = 10^{10.5} - 10^{11.0} \, {\rm M_{\odot}}$ mass bins is not in quantitative agreement with the observed sSFRs, besides the non uni-modal forms \footnote{Our results for the Shark semi-analytic model are in agreement with the findings of \citet{Bravo2020} who showed that Shark reproduces well the color bi-modality, though the transition from predominantly star-forming galaxies to passive objects happens at stellar masses that are larger than the ones found in observations.}. We note that both Shark and L-galaxies allow gas-poor galaxies to be less efficient in SF compared to gas-rich galaxies at a fixed gas surface density, either by using molecular gas-based SF prescriptions or a gas surface density threshold \citep{Lagos2011s} and this can enhance the sSFR bi-modality.

For both SAMs and cosmological simulations a thoughtful tuning of the parameters of the models that describe key processes of galaxy formation is required using observational constraints \citep{Lacey2016}, like the GSMF or the CSFRD. Because of their little computational expense, the tuning of parameters in SAMs can be easily done via MCMC, emulators or genetic algorithms \citep[e.g.][]{Bower2012,Henriques2013,Ruiz2015}, while this task is much more demanding in hydrodynamical simulations. This may be playing an important role in why SAMs perform qualitatively better than most hydrodynamical simulations explored here. However, the tuning done for any case scenario alone cannot guarantee a success for a model since any ``mistakes'' in the recipes that describe gas, star formation and feedback can be compensated through tuning and various observations can be ``successfully'' reproduced. We suggest that the sSFRF can provide a useful test/constraint since it encloses information about galaxy quenching and the relation between SFR and stellar mass. 

\section{Conclusions}
\label{con}

In this work we investigated the Specific Star Formation Rate Function (sSFRF) across different mass scales (${\rm M_{\star}} = 10^{9.0} - 10^{9.5} \, {\rm M_{\odot}}$, ${\rm M_{\star}} = 10^{10} - 10^{10.5} \, {\rm M_{\odot}}$, ${\rm M_{\star}} = 10^{10.5} - 10^{11} \, {\rm M_{\odot}}$ and ${\rm M_{\star}} = 10^{11} - 10^{11.5} \, {\rm M_{\odot}}$) in the Sloan Digital Sky Survey Data Release 7 (SDSS) for objects within the survey's SFRs and Stellar mass confidence limits  \citep[${\rm M_{\star}} > 10^{9} $ ${\rm M_{\odot}}$, ${\rm SFR}$ $>$ $10^{-1.5}$ ${\rm M_{\odot} / {\rm yr}}$, ][]{Weigel2016,Zhaoka2020}. The above enable us to study qualitatively and quantitatively quenching, the distribution of passive/star-forming galaxies and compare the results with the predictions from state-of-the-art cosmological simulations and semi-analytic models (SAMs). The latter have been found to reproduce the observed SFRF and GSMF at $z \sim 0$, while they typically do not suffer from limitations of resolution within the ranges we adopt \citep{Katsianis2017,Zhaoka2020}. Our main conclusions are summarized as follows:

\begin{itemize}

\item The sSFRF (star-forming + quiescent) at different mass scales is bi-modal  (section \ref{SFRSSDSS}). This form is in agreement with the findings of other studies which employed other data-sets at various redshifts \citep{Santini2009,Tzanavaris2010,Lenkic2016,dave2019} and previous studies of the SDSS \citep{Wetzel2011}. Galaxies, within our ${\rm M_{\star}}$ and SFR limits, become more passive with increasing stellar mass but the transition across different mass scales is rather smooth. The low mass end (${\rm M_{\star}} = 10^{9.5} - 10^{10} \, {\rm M_{\odot}}$) is mostly dominated by star forming galaxies with our results from SDSS being in excellent agreement with the main-sequence (star-forming galaxies only) study of \citet{Ilbert2015}. However, moving to the intermediate mass bins (${\rm M_{\star}} = 10^{10} - 10^{10.5} \, {\rm M_{\odot}}$ and ${\rm M_{\star}} = 10^{10.5} - 10^{11} \, {\rm M_{\odot}}$) we see already an arising passive population, which gives a bi-modal form to the sSFRF. At higher masses (${\rm M_{\star}} = 10^{11} - 10^{11.5} \, {\rm M_{\odot}}$) we find a notable star forming population, in agreement with the main-sequence study of \citet{Ilbert2015} and the predictions from cosmological simulations \citep{Dave2017}.

\item The Simba, L-galaxies and Shark models generate bi-modal sSFRFs and thus are qualitatively in agreement with our findings for SDSS (section \ref{simsbina}). We stress that none of these models was tuned to reproduce this behavior. The above is achieved since all models are able to create both star-forming galaxies (sSFR $>$ $10^{-11} {\rm yr^{-1}}$) and an arising with mass passive population (sSFR $<$ $10^{-11} {\rm yr^{-1}}$), within the same ${\rm M_{\star}}$ and SFR limits as those adopted in observations. We note however, that besides that Simba sSFRF is bi-modal \citep{dave2019} it does not match quantitatively SDSS which has typically higher number density values. Simba achieved its sSFR distributions by moving numerous star-forming galaxies towards the passive population via its efficient AGN feedback and as a result underpredicts the peak of the active objects with respect observations. On the other hand, the Shark and L-galaxies semi-analytic models appear to retain the high numbers of their star-forming galaxies and allowed passive objects to maintain low levels of SFR. The presence of the latter generate an adequate quenched population with ${\rm sSFR} \sim 10^{-12} - 10^{-11} \, {\rm yr^{-1}}$. A tuning of both effects found in Simba (decreasing the star forming population to generate more quenched objects) and L-Galaxies (allowing ``dead'' galaxies with SFR = 0 to retain a higher level of SF activity) could bring in better agreement the observed and simulated sSFRFs. 
  
\item Illustris, EAGLE, Mufasa and IllustrisTNG successfully produce the sSFRF of star forming galaxies and are in excellent agreement with the sSFRF from SDSS at sSFR $>$ $10^{-11} {\rm yr^{-1}}$ at all mass regimes. However, the same models are unable to generate an adequate population of passive galaxies with sSFRs between $10^{-11} {\rm yr^{-1}}$ and $10^{-13} {\rm yr^{-1}}$ in order to establish a bi-modality at any mass range considered in this work. Instead the above simulations demonstrate a uni-modal sSFRF in contrast with Simba, L-Galaxies and our results from SDSS. We note that the Illustris, Mufasa and TNG cosmological simulations adopt different feedback schemes, resolutions and volumes but still produce similar results for the sSFRF.  Different models can achieve similar star formation histories (SFHs)\footnote{Besides the fact that different models produce similar SFHs, this can be done by galaxies that are otherwise different, e.g. in terms of inter-stellar medium and circum-galactic medium histories or gas properties \citep{Mitchell2018}. since the parameters of the subgrid physics are constrained following similar methods (e.g. parameter tuning via the GSMF or CSFRD).

\item  We suggest that the sSFRF encloses important information for galaxy quenching and the SFR-${\rm M_{\star}}$ relation. Thus, it can be used as an additional observational constraint/test for cosmological models.} We propose that the feedback and star formation prescriptions that are used in state-of-the-art cosmological simulations have to be reconsidered, involve recipes that allow ``quenched galaxies'' to retain a small level of SF activity (sSFR $=$ ${\rm 10^{-11} {\rm yr^{-1}}}$-${\rm 10^{-12} {\rm yr^{-1}}}$) and generate an adequate passive population/bi-modality even at intermediate masses (${\rm M_{\star}}  = 10^{10} - 10^{10.5} \, {\rm M_{\odot}}$).
  
\end{itemize}

\section*{Acknowledgments}

The authors would like to thank the anonymous referee for their suggestions and comments. These have improved significantly our work. A.K has been supported by the Tsung-Dao Lee Institute Fellowship and Shanghai Jiao Tong University. X.Y. is supported by the national science foundation of China (grant Nos. 11833005, 11890692,11621303) and Shanghai Natural Science Foundation, grant No.15ZR1446700. We also thank the support of the Key Laboratory for Particle Physics, Astrophysics and Cosmology, Ministry of Education.  W.C. acknowledges support from the European Research Council under grant number 670193. C.L. has received funding from the ARC Centre of Excellencefor All Sky Astrophysics in 3 Dimensions (ASTRO 3D), through project number CE170100013. Cosmic Dawn Centre is funded by the Danish National Research Foundation. XZZ acknowledges the supports from  the National Key Research and Development Program of China (2017YFA0402703), the National Science Foundation of China (11773076, 12073078), and the Chinese Academy of Sciences (CAS) through a China-Chile Joint Research Fund (CCJRF no: 1809) administered by the CAS South America Centre for Astronomy (CASSACA).

\section*{Data Availability Statement}

No new data were generated or analyzed in support of this research.

\bibliographystyle{mn2e}	
\bibliography{Katsianis_mnrasRev2.bbl}

\label{lastpage}
\end{document}